\documentclass[numbers, times, sort&compress, 5p]{elsarticle}

\usepackage{mymacros}

\begin{document}

\title{Single flavour optimisations to Hybrid Monte Carlo}
\author[uofa]{Taylor Haar\corref{cor1}}
\ead{taylor.haar@adelaide.edu.au}
\author[uofa]{Waseem Kamleh}
\author[uofa]{James Zanotti}
\author[yoshi]{Yoshifumi Nakamura}

\address[uofa]{CSSM, Department of Physics, The University of Adelaide, Adelaide, SA, Australia 5005}
\address[yoshi]{RIKEN Center for Computational Science, Kobe, Hyogo 650-0047, Japan}

\cortext[cor1]{Corresponding author}

\begin{abstract}
It has become increasingly important to include one or more individual flavours of dynamical fermion in lattice QCD simulations.
This is due in part to the advent of QCD+QED calculations, where isospin symmetry breaking means that the up, down, and strange quarks must be treated separately. 
These single-flavour pseudofermions are typically implemented as rational approximations to the inverse of the fermion matrix, using the technique known as Rational Hybrid Monte Carlo (RHMC).
Over the years, a wide range of methods have been developed for accelerating simulations of two degenerate flavours of pseudofermion, while there are comparatively fewer such techniques for single-flavour pseudofermions.
Here, we investigate two different filtering methods that can be applied to RHMC for simulating single-flavour pseudofermions, namely polynomial filtering (PF-RHMC), and filtering via truncations of the ordered product (tRHMC).
A novel integration step-size tuning technique based on the characteristic scale is also introduced.
Studies are performed on two different lattice volumes, demonstrating that one can achieve significant reductions in the computational cost of single-flavour simulations with these filtering techniques.
\end{abstract}

\begin{keyword}
12.38.Gc \sep Hybrid Monte Carlo algorithm \sep Multiple time scale integration
\sep Single-flavour simulations
\end{keyword}

\maketitle

\thispagestyle{fancy}
\renewcommand{\headrule}{}
\rhead{ADP-18-13/T1061}
\lfoot{\vspace{20pt} \textit{\footnotesize \copyright\ 2018. Distributed under the CC-BY-NC-ND 4.0 license \url{http://creativecommons.org/licenses/by-nc-nd/4.0/.}}} 

\section{Introduction}
\tikzsetfigurename{figure_1.}

Through advances in computational power, the lattice QCD community has achieved the long term goal of performing dynamical simulations at or near the physical point, see e.g.~\cite{Aoki:2009ix, BMW:2010aw, Blum:2014tka, Ishikawa:2015rho, Abdel-Rehim:2015pwa}, using techniques based on the Hybrid Monte Carlo (HMC) algorithm~\cite{Duane:1987}. Nonetheless, the task of generating gauge fields with light quarks remains numerically intensive, requiring access to primary tier supercomputing resources.

The main computational cost in generating lattice configurations comes from evaluating the contribution from the fermion determinant.
This is normally done through the introduction of auxiliary bosonic fields, called pseudofermions, such that the fermion determinant can be expressed as
\begin{equation}
	\det M_f = \int D\phi_f \,D\phi_f^\dag\, \exp \left( - \phi^\dag_f M_f^{-1} \phi_f \right). \label{eq:det_pseudo}
\end{equation}
The key here is that the pseudofermion fields are just complex numbers, so the integral can be evaluated by stochastic means.
A subtlety is that for the integrand to be interpreted as a probability distribution, the fermion matrix must be positive semi-definite, i.e. $\phi_f^\dag M_f^{-1} \phi_f \geq 0$.
In general, the fermion matrix $M_f$ has a complex spectrum, so positive semi-definiteness is not guaranteed.
However, we know that the determinant is real, so for two degenerate fermion flavours, the problem is easily circumvented by combining them into a single integral
\begin{equation}
	(\det M)^2 = \det (M^\dag M) = \int D\phi D\phi^\dag \exp \left( -\phi^\dag (M^\dag M)^{-1} \phi \right).
\end{equation} 
Thus, we can easily perform simulations with the two-flavour pseudofermion action
\begin{equation}
	S_{2f} = \phi^\dag K^{-1} \phi \label{eq:2f_action}
\end{equation}
because $K = M^\dag M$ is positive semi-definite by construction.

In order to include the contribution from a single fermion flavour, another approach must be used.
The most common is Rational HMC (RHMC)~\cite{Kennedy:1998cu}.
First note that for real and positive $\det M$, we have
\begin{equation}
	\det M = \sqrt{\det M^\dag M} = \int D\phi D\phi^\dag \exp \left( - \phi^\dag (M^\dag M)^{-1/2} \phi \right).
\end{equation}
To evaluate this integral in practice, we then replace the matrix inverse by a rational polynomial approximation $R(K) \approx K^{-1/2}$.
Such an approximation is positive semi-definite by construction, so we can use the pseudofermion action
\begin{equation}
	S_{1f} = \phi^\dag R(K) \phi \label{eq:rhmc_action}.
\end{equation}
Note that this is more expensive to simulate with than the two-flavour action \eqref{eq:2f_action} because it is more expensive to calculate $R(K)$ than to invert $K$.

Simulations with pure QCD almost always take advantage of the approximate isospin symmetry present in nature and set the up and the down quarks to be degenerate.
Hence, in a typical $n_f = 2 + 1$ flavour simulation, only the strange quark is handled by RHMC.
In the past, improvements to the RHMC algorithm were not so crucial, as
the relative cost of adding a heavy strange quark to a simulation of light up and down quarks was small.
However, a large range of improvements have been applied to the two-flavour pseudofermion action \eqref{eq:2f_action} to reduce the cost~\cite{Hasenbusch:2001ne, Peardon:2002wb, Luscher:2005, Urbach:2005ji}, such that in a modern simulation the relative cost of including the strange quark can be significant.

Moreover, advances in lattice QCD calculations of certain quantities have reached the point that some collaborations are incorporating dynamical QED effects into their gauge field configurations~\cite{Aoki:2012st, BMW:2013, Horsley:2015vla, Horsley:2015eaa}, for the precision reached for these quantities has made electromagnetic effects significant.
In such simulations, isospin symmetry is explicitly broken as the up and down quarks have different charges $q_u \neq q_d$.
This necessitates the use of $n_f = 1 + 1 + 1$ simulations at or near the physical point.
This further motivates the development of modifications to the single-flavour RHMC algorithm to parallel improvements made in the degenerate two-flavour case.

\section{Method}
\tikzsetfigurename{figure_2.}

\subsection{Hybrid Monte Carlo (HMC)}
In Lattice QCD, we want to sample gauge fields $U$ and pseudofermion fields $\phi$ from the distribution $\frac{1}{Z} \exp(-S[U, \phi])$, where the lattice action $S = S_G + S_F$ is the sum of the gauge action $S_G[U]$ and the pseudofermion action $S_F[U,\phi]$.
The usual way we achieve this is with Hybrid Monte Carlo (HMC)~\cite{Duane:1987}.

At the start of each molecular dynamics trajectory, we generate appropriately distributed pseudofermion fields $\phi$ from random number fields $\chi,$ drawn from a Gaussian heatbath $\rho(\chi) \sim e^{-\chi^\dag \chi}$.
Then, to generate $U$, we introduce a fictitious momentum field $P$ conjugate to the gauge field and construct the Hamiltonian
\begin{equation}
	H[P,U] = \sum \tr{[P^2]} + S[U,\phi].
\end{equation}
The system is updated via coupled molecular dynamics integration steps,
\begin{align*}
	\hat{T}[\epsilon]:\quad & (P,U) \rightarrow (P, e^{ih P}U) \\
	\mathrm{and} \quad \hat{S}[\epsilon]:\quad & (P,U) \rightarrow (P - h F, U)
\end{align*}
with force term $F \equiv \pdiff{S}{U}$ and step-size $h$, such that the Hamiltonian $H$
is approximately preserved via Hamilton's equations.
A full molecular dynamics trajectory, typically of length $\tau = 1$ in simulation time, is used to evolve the system from
$(P,U)$ to a new candidate state $(P',U')$.
This new state then undergoes a Metropolis acceptance step, where the probability of acceptance is
\begin{equation}
	P_{acc} = \min \left[1, \exp \left( H[P,U] - H[P',U'] \right) \right].
\end{equation}
This last step ensures that the method is exact, such that the Markov chain of gauge configurations $U^{(0)} \rightarrow U^{(1)} \rightarrow U^{(2)} \ldots$\ tends to the desired distribution $\frac{1}{Z} \exp(-S[U,\phi])$.

The cost of generating configurations via HMC is predominantly due to the evaluation of the force term $F = \pdiff{S}{U}$ at each integration step.
In the case of RHMC, the fermion force term involves inverses of the matrix $K$ plus a linear shift (see \ref{app:forces}), so the cost of inverting $K$, and hence $M$, plays a major role in the overall computational cost.

\subsection{Filtering methods}

As we move to lighter quark masses, the cost of generating configurations goes up in two ways.
First, the smaller mass causes the fermion matrix $M$ to become more singular, which increases the computational effort required to invert it.
Second, light masses increase the chance that the molecular dynamics integration will become unstable,
so the integration step-size needs to be reduced to maintain a good acceptance rate.
Thus, to effectively simulate at light quark masses, we need methods that tackle both of these problems.

By enabling separate treatment of the high and low frequency
dynamics, filtering methods provide an effective means to ameliorate
the cost of simulating at light quark masses.  Noting that for
matrices $L, F$ invertible,
\begin{equation}
	\det L = \frac{\det [FL]}{\det F},
\end{equation}
a filtering method separates a pseudofermion action
\begin{equation}
	S = \phi^\dag L^{-1}\, \phi
\end{equation}
into multiple terms,
\begin{equation}
	S_{\mathrm{filtered}} = \phi_1^\dag F \phi_1 + \phi_2^\dag F^{-1} L^{-1}\, \phi_2. \label{eq:filter}
\end{equation}
We say that $F$ acts as a filter for the matrix kernel $L^{-1}$.
A good choice of filter ensures that $F$ acts as a preconditioner for $L,$ such that $F^{-1}L^{-1}$ has a reduced force term.
Secondly, we require that the filter term has a relatively cheap force to evaluate and captures the high energy modes of the system. Then we can use multiple time-scales~\cite{Sexton:1992,Peardon:2002wb} to reduce the number of integration steps for the relatively expensive correction term.

\subsection{Polynomial filtering}

Polynomial filtering (PF)~\cite{Kamleh:2011dc} uses a polynomial filter $P$ that approximates the matrix kernel $L^{-1}.$ For a single-flavour pseudofermion \eqref{eq:rhmc_action}, we choose an approximation $P(K) \approx K^{-1/2}$, giving the action
\begin{equation}
	S_{PF-RHMC} = \phi_1^\dag P(K) \phi_1 + \phi_2^\dag P(K)^{-1} R(K) \phi_2. \label{eq:pfrhmc}
\end{equation}

\begin{figure}[t]
\centering
\begin{tikzpicture}[trim axis left]

\begin{axis}[
	cycle list name=mstone_d,
	xmin=0, xmax=3,
	ymin=0, ymax=1.5,
	xlabel=$z$,
	ylabel=$\sqrt{z} P(z)$,
	legend pos=south east,
	]

\renewcommand{\matrixopfile}{\figdir/data/PFRHMC/pfrhmc_approx_poly.txt}

\draw[thin, gray]
	(0,1) -- (3,1);

\foreach \i in {1,2,3} {
\addplot+[mark=none]
	table[
		x=x,
		y=y\i,
	]
	{\matrixopfile};
}

\legend{$p=4$, $p=8$, $p=12$}

	]

\end{axis}

\end{tikzpicture}
\caption{$\sqrt{z} P(z)$ for various Chebyshev polynomials $P(z)$ with order $p$ and range $[10^{-4}, 3]$. Closeness to unity indicates how well $P(z)$ approximates $z^{-1/2}$. \label{fig:pfrhmc_filter}}
\end{figure}
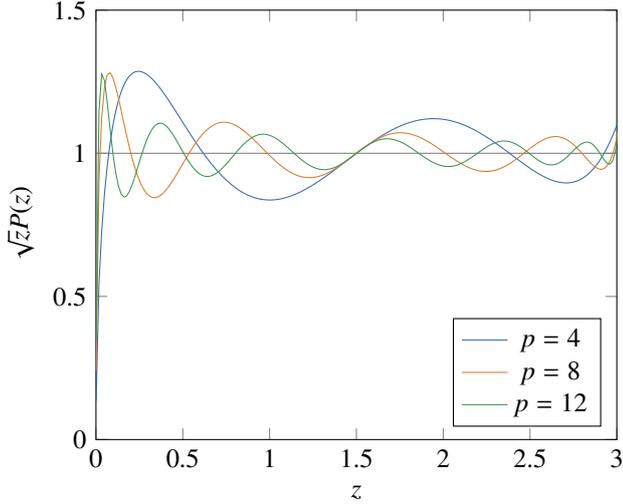

Consider our requirements for a good filtering method.
The polynomial term $S_1 = \phi_1^\dag P(K) \phi_1$ has a force which is easy to calculate as there is no need to invert $K$ (see \ref{app:forces}), and captures the higher energy modes because it is an approximation to $K^{-1/2}$.
Figure~\ref{fig:pfrhmc_filter} demonstrates how good this approximation is for various polynomial filters with Chebyshev polynomials.

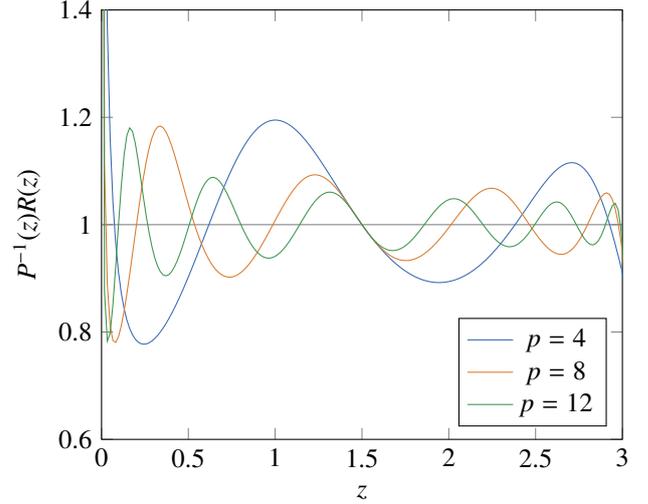
\begin{figure}[t]
\centering
\begin{tikzpicture}[trim axis left]

\begin{axis}[
	xmin=0, xmax=3,
	ymin=0.6, ymax=1.4,
	cycle list name=mstone_d,
	xlabel=$z$,
	ylabel=$P^{-1}(z) R(z)$,
	legend pos=south east,
	]

\renewcommand{\matrixopfile}{\figdir/data/PFRHMC/pfrhmc_approx_corr.txt}

\draw[thin, gray]
	(0,1) -- (3,1);

\foreach \i in {1,2,3} {
\addplot+[mark=none]
	table[
		x=x,
		y=y\i,
	]
	{\matrixopfile};
}

\legend{$p=4$, $p=8$, $p=12$}

\end{axis}

\end{tikzpicture}
\caption{$P^{-1}(z) R(z)$ for various Chebyshev polynomials $P(z)$ with order $p$ and range $[10^{-4}, 3]$. $R(z)$ is the 20th order Zolotarev approximation in Table~\ref{tbl:zolo_eg}. \label{fig:pfrhmc_correction}}
\end{figure}

Now consider the filtered term, $\phi_2^\dag P(K)^{-1} R(K) \phi_2.$
So long as the roots of $P(K)$ are not too close to zero, there is little additional expense to calculate this term via a multi-shift solver.
This term also approximates unity (see Figure \ref{fig:pfrhmc_correction} for some examples) which leads to a small force term and thus permits a coarser integration step-size.
Therefore, both terms satisfy the requirements for a good filter, and the computational cost can be reduced by setting $h_1 < h_2$.

\subsection{Filtering via truncations of the ordered product (tRHMC)}
Another filtering method (used by e.g.~\cite{Luscher:2012av}), which is specific to RHMC, starts from the ordered product expansion of the rational approximation:
\begin{equation}
	R(K) = c_n \prod_{k=1}^n \frac{K + a_k}{K + b_k}, \label{eq:ratapp}
\end{equation}
where $n$ is the order of the approximation and $a_k, b_k, c_n$ are real numbers with the factors strictly ordered such that $a_k > a_{k+1}$, $b_k > b_{k+1}$.
From this form, we can define a partial factor as
\begin{equation}
	R_{i,j}(K) = (c_n)^{\delta_{i,0}}  \prod_{k=i+1}^j \frac{K + a_k}{K + b_k},
\end{equation}
noting that the normalisation $c_n$ is only included if $i=0$.
A useful property of the Zolotarev rational approximation we use in this work is that the partial factor $R_{0,t}(K)$ with $t < n$ also forms an approximation to $K^{-1/2}$.
Hence, we can use this truncation of the ordered product as a filter,
\begin{equation}
	S_{tRHMC} = \phi_1^\dag R_{0,t}(K) \phi_1 + \phi_2^\dag R_{t,n}(K) \phi_2. \label{eq:trhmc_action}
\end{equation}
We refer to this method as \emph{truncated ordered product (top)} filtering, or simply tRHMC for short.
Note that both terms here are rational functions of $K$, which makes implementation in existing RHMC code very easy.

In order for tRHMC to be a good filtering method, we require $R_{0,t}(K)$ to have a cheap force term and to form a good approximation to $K^{-1/2}$ at high energies.
The first criteria is satisfied because the leftmost shifts in the product form are relatively large due to the shift ordering.
To demonstrate the second criteria, we consider the case of a 20th order Zolotarev approximation with range $[5 \times 10^{-5}, 3]$, whose coefficients are given in Table~\ref{tbl:zolo_eg}.
Figure~\ref{fig:trhmc_filter} shows how closely the truncated term $R_{0,t}(K)$ approximates $K^{-1/2}$ by plotting $z R_{0,t}(z) \approx 1$.
One can see that all the shown truncations produce a good approximation at large $K$, and get better at small $K$ for larger truncation orders $t$. Thus, the truncated term captures the high energy molecular dynamics of the system, and thus behaves as a good filter.

\begin{figure}[t]
\centering
\begin{tikzpicture}[trim axis left]

\begin{axis}[
	cycle list name=mstone_d,
	xmin=0, xmax=3,
	ymin=0, ymax=1.2,
	xlabel=$z$,
	ylabel=$\sqrt{z} R_{0,t}(z)$,
	legend pos=south east,
	]

\renewcommand{\matrixopfile}{\figdir/data/tRHMC/trhmc_approx_trunc.txt}

\draw[thin, gray]
	(0,1) -- (3,1);

\foreach \i in {1,2,3,4} {
\addplot+[mark=none]
	table[
		x=x,
		y=y\i,
	]
	{\matrixopfile};
}

\legend{$t=4$, $t=6$, $t=8$, $t=10$}

	]

\end{axis}

\end{tikzpicture}
\caption{$\sqrt{z} R_{0,t}(z)$ for various truncation orders $t$. $R(z)$ is the 20th order Zolotarev approximation given in Table \ref{tbl:zolo_eg}. Closeness to unity indicates how well $R_{0,t}(z)$ approximates $z^{-1/2}$. \label{fig:trhmc_filter}}
\end{figure}
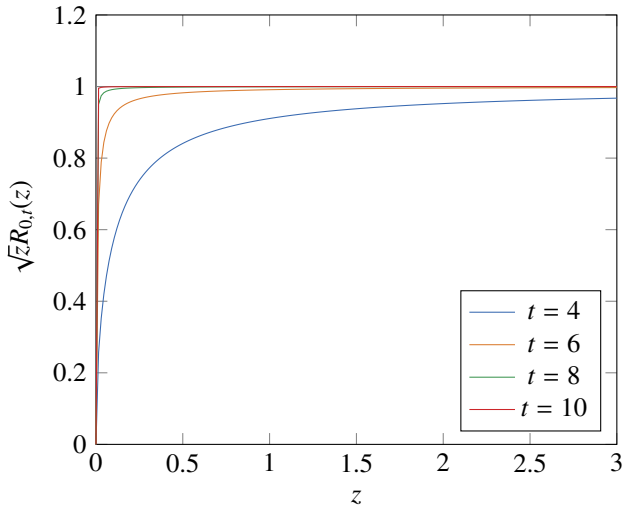

\begin{figure}[t]
\centering
\begin{tikzpicture}[trim axis left]

\begin{axis}[
	xmin=0, xmax=3,
	ymin=0.9, ymax=1.1,
	restrict y to domain=0.9:2,
	cycle list name=mstone_d,
	xlabel=$z$,
	ylabel=$R_{t,n}(z)$,
	legend pos=south east,
	]

\renewcommand{\matrixopfile}{\figdir/data/tRHMC/trhmc_approx_corr.txt}

\draw[thin, gray]
	(0,1) -- (3,1);

\foreach \i in {1,2,3,4} {
\addplot+[mark=none]
	table[
		x=x,
		y=y\i,
	]
	{\matrixopfile};
}

\legend{$t=4$, $t=6$, $t=8$, $t=10$}

\end{axis}

\end{tikzpicture}
\caption{$R_{t,n}(z)$ for various truncation orders $t$. $R(z)$ is the 20th order Zolotarev approximation given in Table \ref{tbl:zolo_eg}. The filtered term rapidly approaches unity as the truncation order $t$ increases. \label{fig:trhmc_correction}}
\end{figure}
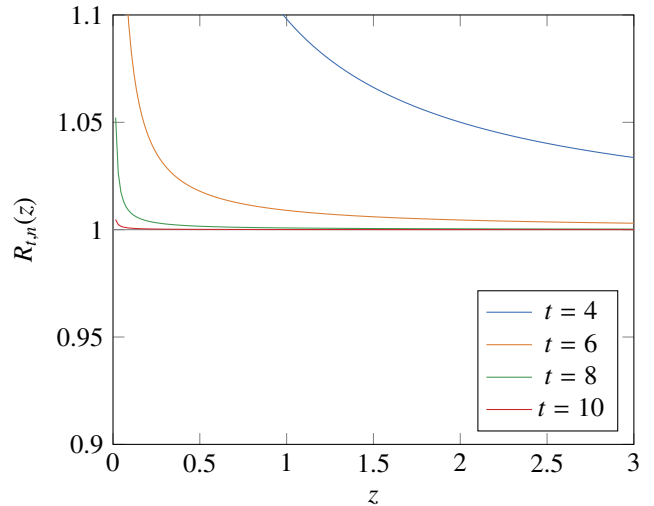

Now consider the correction term.
$R_{t,n}(K)$ is no more expensive to calculate than the full approximation $R(K) = R_{0,n}(K)$ when using a multi-shift solver.
This term should also approximate unity by construction.
To show this, we plot $R_{t,n}(K)$ for various truncation orders using our example Zolotarev rational approximation in Figure \ref{fig:trhmc_correction}.
We can see that the correction term approximates unity more and more closely as we increase the truncation order $t$.
Closeness to unity indicates a smaller molecular dynamics force, which enables a coarser integration step-size to be used during HMC.
This, along with the properties of the truncated term, shows that tRHMC satisfies the requirements for a good filtering method.

It is important to note that the tRHMC filtering method is distinct from splitting the sum-over-poles form of $R(K)$, with different poles being placed on different time scales (used by e.g.~\cite{Allton:2007hx}).
The pole-splitting method distributes the poles across time-scales based simply on inversion cost, and lacks the additional benefits of the truncated term $R_{0,t}(K)$ in tRHMC; it approximates $K^{-1/2}$ and hence also acts as a high pass filter.

Another technique for improving the performance of RHMC is the n\textsuperscript{th} root trick~\cite{Clark:2006}, where some number $n$ of RHMC pseudofermions with approximations $R(K) \approx K^{-1/2n}$ are used to simulate a single fermion flavour.
In comparison to this, a filtering method such as tRHMC allows for splitting the energy modes of the action such that the most expensive parts are not evaluated as often, potentially enabling a larger cost reduction.
Our tRHMC filtering method may be also more efficient on some hardware architectures due to the memory bandwidth benefits of evaluating multiple small order rational polynomial terms, rather than a single term of large order.

\section{Results}
\tikzsetfigurename{figure_3.}

We begin by performing comparisons on a $16^3 \times 32$ lattice with two (degenerate) individual flavours of Wilson fermion, i.e. a $n_f = 2$ simulation performed with $n_f = 1 + 1$ methods.
This lattice has pion mass $m_{\pi} \sim 400$~MeV and lattice spacing $\sim 0.08$~fm.
This lattice was chosen in order to make it easy to compare the relative cost with the degenerate $n_f = 2$ methods analysed in our previous work~\cite{Haar:2016bwe}.
Note that while the two single-flavour fermion actions are treated equally on this lattice, this is not a requirement for the filtering methods studied here: these apply to each fermion flavour individually, and hence allow for the possibility that the fermions have different masses and/or electric charges.

The cost function we use to compare the performance of our methods is the same as in previous work~\cite{Haar:2016bwe}, namely
\begin{equation}
	C = N_{mat}/P_{acc} \label{eq:cost}
\end{equation}
where $N_{mat}$ is the total number of fermion matrix $M$ multiplications required per trajectory and $P_{acc}$ is the average Metropolis acceptance rate.
This approximates the cost of generating uncorrelated gauge configurations.
For the $16^3 \times 32$ lattice, our results are an average over $\sim 2000$ trajectories per set of filter parameters.

\subsection{RHMC}

The baseline for comparing our filtering schemes is obviously the cost of a standard RHMC simulation.
This has pseudofermion action
\begin{equation}
	S_{RHMC}[U,\phi] =  \phi^\dag R(K) \phi.
\end{equation}
In all our simulation results that follow, it is implicitly assumed that there are two copies of the pseudofermion action, one for each degenerate flavour, with independent pseudofermion fields.
The rational approximation $R(K)$ we use for the $16^3 \times 32$ lattice is a 20th order Zolotarev approximation on the interval $[5 \times 10^{-5}, 3],$ refer to Table~\ref{tbl:zolo_eg} for the coefficients.
The eigenvalue range for this lattice is $[\lambda_{\mathrm{min}}, \lambda_{\mathrm{max}}] = [6.8(1) \times 10^{-5}, 2.18949(5)]$.
We use the Zolotarev rational approximation in this work as it is optimal for approximating the inverse square root $K^{-1/2}$ (see e.g.~\cite{Chiu:2002eh}).

We choose the rational approximation here with a range that encompasses the eigenvalue range of the fermion matrix in question, and use a sufficiently high order to ensure good precision as required for the Metropolis acceptance step.
However, a rational approximation does not need a very high order for it to get close to floating-point precision ($n=20$ here).

With the rational approximation fixed, there are only two parameters to tune for plain RHMC:
these are the step-sizes $\{h_0, h_1\}$, corresponding to the gluon and fermion actions respectively.
It is often useful to express these as step counts $n_i = \tau / h_i$, where $\tau$ ($= 1$ here) is the trajectory length.
We choose to fix the gluon action's step-size to $n_0 = n_G = 480$ throughout this paper, because $S_G$ is very cheap to calculate compared to any of the fermion terms we encounter.
This leaves just one parameter, making tuning quite simple.

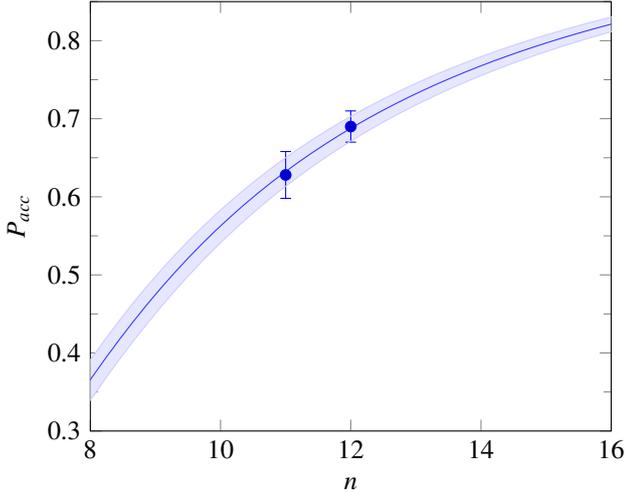
\begin{figure}[t]
\centering
\begin{tikzpicture}[trim axis left]

\begin{axis}[
	xlabel=$n$,
	ylabel=$P_{acc}$,
	y errors,
	xmin=8, xmax=16,
	ymin=0.3, ymax=0.85,
	minor y tick num=1,
]

\addplot table [
	x=N_md, 
	y=Pacc,
	y error=Pacc_err,
	only marks,
	]
	{\figdir/data/RHMC/rhmc_erfc_data.txt};

\addplot+[
	band_1_cen, 
	]
	table [
	x=n,
	y=P_fit,
	mark=none,
	]
	{\figdir/data/RHMC/erfc_fit_data.txt};

\addplot+[
	name path=low,
	band_1_out,
	]
	table [
	x=n,
	y=P_low,
	mark=none,
	]
	{\figdir/data/RHMC/erfc_fit_data.txt};

\addplot+[
	name path=high,
	band_1_out,
	]
	table [
	x=n,
	y=P_high,
	mark=none,
	]
	{\figdir/data/RHMC/erfc_fit_data.txt};
	
\addplot+[
	band_1,
	]
	fill between[
		of=low and high,
		soft clip={domain=8:16},
	];


\end{axis}

\end{tikzpicture}
\caption{Fit of RHMC $P_{acc}$ data on the $16^3 \times 32$ lattice to the complementary error function \eqref{eq:pacc_erfc}.
The fit is shown as a faded band, and has characteristic scale  $c = 0.156(4)$. \label{fig:rhmc_erfc}}
\end{figure}

We target an acceptance rate $P_{acc}$ in the range $0.65 - 0.75$, as this provides a good balance between accepting candidate configurations and the cost of generating uncorrelated samples.
To achieve this, we note that in pure HMC, the acceptance rate as a function of the step-size $h$ can be modelled by
\begin{equation}
	P_{acc} = \mathrm{erfc} \left( \frac{c^2}{h^2} \right) \label{eq:pacc_erfc}
\end{equation}
where $c$ is a fitting parameter known as the characteristic scale.
In the case of plain RHMC, the rational term controls the majority of the dynamics, so to a good approximation we can use the fermion step-size $h_1$ in this expression.
We thus evaluate $P_{acc}$ for a variety of $h_1$ values, then fit to \eqref{eq:pacc_erfc} to find the step-size $h_1$ that gives the desired $P_{acc}$.
The fit for our configuration is shown in Figure \ref{fig:rhmc_erfc}, and from this we choose the step-count $n_1 = 12$.

Using the step counts $(n_1, n_0) = (12, 480)$, we find that the number of matrix operations required per trajectory for pure RHMC is $N_{mat} = 46,960 \pm 120$
with an acceptance rate of $P_{acc} = 0.69(1)$.
This gives a normalised cost of
\begin{equation}
	C_{RHMC} = 68,100 \pm 1,200, \label{eq:cost_rhmc}
\end{equation}
setting the benchmark that our filtering methods will have to beat.

\subsection{PF-RHMC}

For polynomial filtered RHMC \eqref{eq:pfrhmc}, we generate Chebyshev polynomial approximations $P(K)$ to $K^{-1/2}$ on the same range as the rational approximation.
This leaves a choice of polynomial order $p$, so in order to determine which integer value of $p$ is optimal, we sample a broad set of polynomial orders.

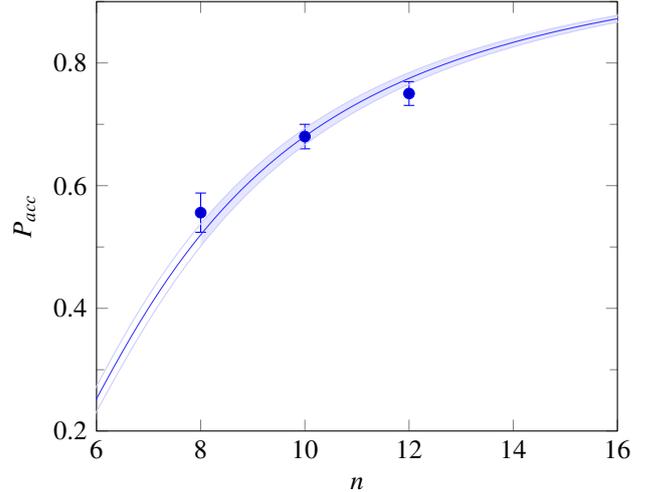
\begin{figure}[t]
\centering
\begin{tikzpicture}[trim axis left]

\begin{axis}[
	xlabel=$n$,
	ylabel=$P_{acc}$,
	y errors,
	xmin=6, xmax=16,
	ymin=0.2, ymax=0.9,
	minor y tick num=1,
]

\addplot table [
	x=N_md, 
	y=Pacc,
	y error=Pacc_err,
	only marks,
	]
	{\figdir/data/PFRHMC/erfc_j135X.txt};

\addplot+[
	band_1_cen, 
	]
	table [
	x=n,
	y=P_fit,
	mark=none,
	]
	{\figdir/data/PFRHMC/erfc_fit_j135X_data.txt};

\addplot+[
	name path=low,
	band_1_out,
	]
	table [
	x=n,
	y=P_low,
	mark=none,
	]
	{\figdir/data/PFRHMC/erfc_fit_j135X_data.txt};

\addplot+[
	name path=high,
	band_1_out,
	]
	table [
	x=n,
	y=P_high,
	mark=none,
	]
	{\figdir/data/PFRHMC/erfc_fit_j135X_data.txt};
	
\addplot+[
	band_1,
	]
	fill between[
		of=low and high,
		soft clip={domain=8:16},
	];


\end{axis}

\end{tikzpicture}
\caption{Fit of PF-RHMC $p=4$ $P_{acc}$ data on the $16^3 \times 32$ lattice to the complementary error function \eqref{eq:pacc_erfc}.
The fit is shown as a faded band, and has characteristic scale $c = 0.185(4)$. \label{fig:pfrhmc_erfc}}
\end{figure}

There are now three step-sizes to tune, $\{h_0, h_1, h_2\}$. As with plain RHMC, we set the gluon action on a fine scale $n_0 = 480$.
To tune the other two step-sizes, we start by fixing the step-size ratio $h_2/h_1$ through \emph{force balancing} the maximal forces $F_i,$ as was done in previous work~\cite{Haar:2016bwe}. In practice, this means we set
\begin{equation}
	F_i h_i \simeq \mathrm{constant}. \label{eq:force_balance}
\end{equation}
This leaves just one step-size to tune: the coarsest one $h_2$, corresponding to the correction term $S_2 = \phi^\dag_2 P(K)^{-1} R(K) \phi_2$.
This can be tuned, as in the RHMC case, by fitting the acceptance rate as a function of $h_2$ to the complementary error function.
As an example, the fit for $p=4$ is shown in Figure \ref{fig:pfrhmc_erfc}, which suggests choosing $n_2 = 10$.
The forces for all polynomial orders are shown in the middle column of Figure \ref{fig:force_1f}, and the resultant step-sizes are shown in Table \ref{tbl:pfrhmc}.

\begin{figure}[tb]
\centering
\begin{tikzpicture}[trim axis group left]

\begin{groupplot}[
	mysmall,
	group style={
		group size=3 by 1,
		horizontal sep=5pt,
		yticklabels at=edge left,
		},
	cycle list name=mstone_d,
	ymin=0, ymax=160000,
	scaled y ticks=base 10:-4,
	ytick scale label code/.code={},
	minor y tick num=1,
	ymajorgrids,
	y errors,
	xlabel absolute group,
	]


\nextgroupplot[
	title=RHMC,
	ylabel=$C / 10^4$,
	only marks,
	xtick=data,
	xticklabels={},
	width=1cm,
	xmin=-0.5, xmax=0.5,
]

\renewcommand{\matrixopfile}{\figdir/data/RHMC/witers_j1350.txt}

\filldraw[draw=band_1_out, fill=band_1]
 (-1, 68060-\autocorr*1170) rectangle (1, 68060+\autocorr*1170);
\draw[band_1_cen_light] (-1, 68060) -- (1, 68060);

\addplot
	table[
		x=ID,
		y=iter_SF,
		y error expr={\thisrow{iter_SF_err}*\autocorr},
	]
	{\matrixopfile};

\pgfplotsset{cycle list shift=1}
\addplot
	table[
		x=ID,
		y=iter_F1,
		y error expr={\thisrow{iter_F1_err}*\autocorr},
	]
	{\matrixopfile};
\pgfplotsset{cycle list shift=0}

\addplot+[red, mark=*]
	table[
		x=ID,
		y=iter_tot,
		y error expr={\thisrow{iter_tot_err}*\autocorr},
	]
	{\matrixopfile};


\nextgroupplot[
	title=PF-RHMC,
	xlabel=$p$,
	only marks,
	width=3cm,
]

\renewcommand{\matrixopfile}{\figdir/data/PFRHMC/witers_j135X.txt}

\filldraw[draw=band_1_out, fill=band_1]
 (2, 68060-\autocorr*1170) rectangle (18, 68060+\autocorr*1170);
\draw[band_1_cen_light] (2, 68060) -- (18, 68060);

\addplot 
	table[
		x=p,
		y=iter_SF,
		y error expr={\thisrow{iter_SF_err}*\autocorr},
	]
	{\matrixopfile};

\foreach \i in {1,2} {
\addplot 
	table[
		x=p,
		y=iter_F\i,
		y error expr={\thisrow{iter_F\i_err}*\autocorr},
	]
	{\matrixopfile};
}

\addplot+[red, mark=*] 
	table[
		x=p,
		y=iter_tot,
		y error expr={\thisrow{iter_tot_err}*\autocorr},
	]
	{\matrixopfile};


\nextgroupplot[
	title=tRHMC,
	xlabel=$t$,
	only marks,
]

\renewcommand{\matrixopfile}{\figdir/data/tRHMC/witers_j138X.txt}

\filldraw[draw=band_1_out, fill=band_1]
 (0, 68060-\autocorr*1170) rectangle (10, 68060+\autocorr*1170);
\draw[band_1_cen_light] (0, 68060) -- (10, 68060);

\addplot
	table[
		x=t,
		y=iter_SF,
		y error expr={\thisrow{iter_SF_err}*\autocorr},
	]
	{\matrixopfile};

\foreach \i in {1,2} {
\addplot
	table[
		x=t,
		y=iter_F\i,
		y error expr={\thisrow{iter_F\i_err}*\autocorr},
	]
	{\matrixopfile};
}

\addplot+[red, mark=*]
	table[
		x=t,
		y=iter_tot,
		y error expr={\thisrow{iter_tot_err}*\autocorr},
	]
	{\matrixopfile};

\end{groupplot}

\end{tikzpicture}
\caption{Cost function for RHMC and the single-filter actions on the $16^3 \times 32$ lattice, using force balancing.
Filled circles are the total cost.
Empty squares are the component of the total cost due to action initialisation.
For RHMC, empty circles are the cost component due to calculating the force term $F$.
Otherwise,
empty triangles are the cost component due to calculating $F_1$,
and empty circles due to calculating $F_2$.
The faded band is the cost of plain RHMC, included for ease of comparison. \label{fig:1f_rhmc_cost}}
\end{figure}
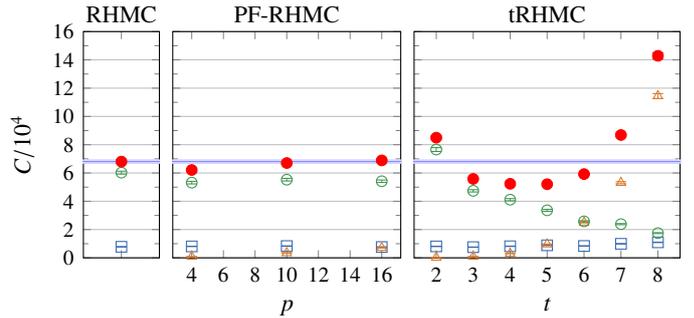

The cost function for PF-RHMC with force-balanced points is shown in the middle plot of Figure~\ref{fig:1f_rhmc_cost}.
The optimal point is at $p=4$ with $C = 62,100 \pm 1,100$, which is only $9\%$ cheaper than plain RHMC \eqref{eq:cost_rhmc}.
The main contributor to this is the cost of evaluating the correction term's forces $F_2$ (empty circles in Figure~\ref{fig:1f_rhmc_cost}), which does not go down as we increase the polynomial order.
The computational cost of each force term $F_2$ evaluation is relatively constant in $p$ due to the use of a multi-shift solver (see \ref{app:forces}). While Figure~\ref{fig:force_1f} shows that the magnitude of $F_2$ does decrease as $p$ increases, the corresponding step count $n_2$ required to maintain an acceptance rate in the range $0.65$--$0.75$ only changes marginally (see Table \ref{tbl:pfrhmc}).

\begin{figure}[tb]
\centering
\begin{tikzpicture}[baseline, trim axis group left]

\renewcommand{\maxforcefile}{\figdir/data/PFRHMC/fmax_j135X.txt}
\renewcommand{\avgforcefile}{\figdir/data/PFRHMC/favg_j135X.txt}
\newcommand{\maxforcefilea}{\figdir/data/tRHMC/fmax_j138X.txt}
\newcommand{\avgforcefilea}{\figdir/data/tRHMC/favg_j138X.txt}
\newcommand{\maxforcefileb}{\figdir/data/RHMC/fmax_j1350.txt}
\newcommand{\avgforcefileb}{\figdir/data/RHMC/favg_j1350.txt}

\pgfplotsset{
	left plot/.style={
		bar width=5pt,
		xtick=data,
		enlarge x limits=0.25,
	},
	right plot/.style={
		bar width=4pt,
		xtick=data,
		enlarge x limits=0.1,
	},
}

\begin{groupplot}[
	mysmall,
	group style={
		group size=3 by 2,
		xlabels at=edge bottom,
		horizontal sep=10pt,
		vertical sep=10pt,
		xticklabels at=edge bottom,
		yticklabels at=edge left,
		},
	ybar,
	xlabel absolute group=-2.5em,
	ylabel absolute group=1.5em,
	]


\nextgroupplot[
	title={RHMC},
	ylabel={$F_1$},
	ymin=0, ymax=6,
	left plot,
	width=1cm,
]

\addplot
	table[
		x=ID,
		y=F_F1,
		y error=F_F1_err,
	]
	{\maxforcefileb};

\addplot
	table[
		x=ID,
		y=F_F1,
		y error=F_F1_err,
	]
	{\avgforcefileb};

\nextgroupplot[
	title={1PF-RHMC},
	ymin=0, ymax=6,
	left plot,
	width=2.5cm,
]

\addplot
	table[
		x=p,
		y=F_F1,
		y error=F_F1_err,
	]
	{\maxforcefile};

\addplot
	table[
		x=p,
		y=F_F1,
		y error=F_F1_err,
	]
	{\avgforcefile};

\nextgroupplot[
	title={1tRHMC},
	ymin=0, ymax=6,
	right plot,
]

\addplot
	table[
		x=t,
		y=F_F1,
		y error=F_F1_err,
	]
	{\maxforcefilea};

\addplot
	table[
		x=t,
		y=F_F1,
		y error=F_F1_err,
	]
	{\avgforcefilea};


\nextgroupplot[
	ylabel={$F_2$},
	ymin=0, ymax=4,
	left plot,
	width=1cm,
]

\addplot+[
	fill=blue!20!white,
	draw=blue!50!white,
]
	table[
		x=ID,
		y=F_F1,
		y error=F_F1_err,
	]
	{\maxforcefileb};

\addplot+[
	fill=red!20!white,
	draw=red!50!white,
]
	table[
		x=ID,
		y=F_F1,
		y error=F_F1_err,
	]
	{\avgforcefileb};

\nextgroupplot[
	xlabel={$p$},
	ymin=0, ymax=4,
	left plot,
	width=2.5cm,
]

\addplot
	table[
		x=p,
		y=F_F2,
		y error=F_F2_err,
	]
	{\maxforcefile};

\addplot
	table[
		x=p,
		y=F_F2,
		y error=F_F2_err,
	]
	{\avgforcefile};

\nextgroupplot[
	xlabel={$t$},
	ymin=0, ymax=4,
	right plot,
]

\addplot
	table[
		x=t,
		y=F_F2,
		y error=F_F2_err,
	]
	{\maxforcefilea};

\addplot
	table[
		x=t,
		y=F_F2,
		y error=F_F2_err,
	]
	{\avgforcefilea};

\end{groupplot}
\end{tikzpicture}
\caption{Forces for the 1-filter actions on the $16^3 \times 32$ lattice. Left-hand bars show the maximal force, while right-hand bars show the average. The single force term for plain RHMC is included for comparison on both force terms. \label{fig:force_1f}}
\end{figure}
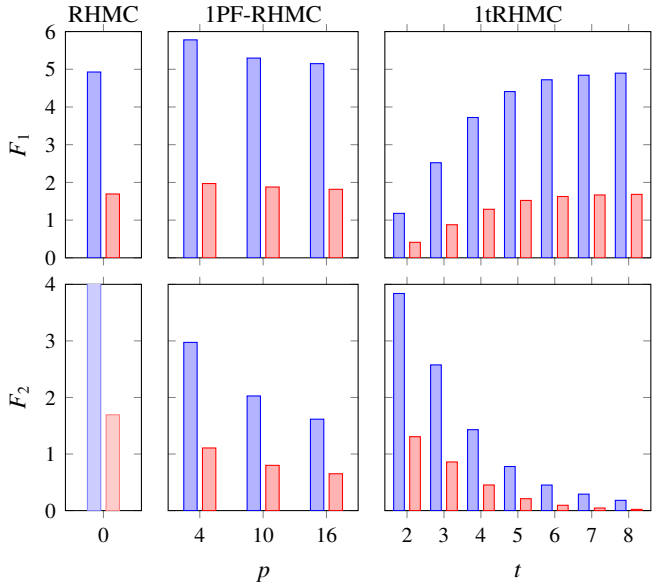

\subsection{tRHMC}
For truncated RHMC \eqref{eq:trhmc_action}, the only novel tunable parameter is the truncation order $t$.
We sample $t$ then tune the step-sizes with force balancing and acceptance rate fitting, just as in the case of PF-RHMC.
The forces are shown in the right-hand column of Figure \ref{fig:force_1f}, and the resultant configuration points are shown in Table \ref{tbl:trhmc}.

The cost function is shown in right-hand plot of Figure \ref{fig:1f_rhmc_cost}, and has a minimum at $t=5$ with $C=52,100 \pm 1,000$.
This is a more significant improvement than PF-RHMC: it is $24 \%$ cheaper than plain RHMC.
However, note that the cost increases dramatically if we choose a truncation too low at $t=2$, or too high at $t=8$.

\subsection{Characteristic scale tuning} \label{sec:cscale}

Simulations of a single fermion flavour inherently have a reduced force relative to the degenerate two-flavour case, simply due to the fact that the fermion matrix is no longer squared.
This implies that the fermion matrix inversions can take place on a much coarser scale for the single-flavour molecular dynamics integration.
As a consequence of this coarse granularity, when applying filtering techniques tuning this scale to maintain an acceptance rate in the desired range can be problematic, as we saw for PF-RHMC and tRHMC above.
In particular, the size of the forces for the correction term for tRHMC can become very small, making the force-balancing method of choosing step sizes highly sub-optimal.
This motivates us to try and refine this approach.

In Figure~\ref{fig:1f_rhmc_cost}, we see that choosing a truncation order $t$ a little too low or high can cause a significant increase in cost. In order to mitigate the effects of choosing a slightly sub-optimal truncation order, we propose a different step-size tuning technique we denote \emph{characteristic scale tuning}, or \emph{c-scale tuning} for short.

The basic premise starts by assuming that the acceptance rate is largely dictated by the coarsest step-size $h_2$, because the correction term $S_2$ carries the long range dynamics and hence the majority of the computational effort.
This term typically has a relatively noisy force term in practice. Hence, the force balancing method described by \eqref{eq:force_balance} might make the step-size for the filter term $h_1$ needlessly small.

The technique of c-scale tuning is an additional step on top of force balancing.
The force-balanced point with suitable acceptance rate is refined by keeping $h_2$ fixed and tuning the next step-size $h_1$ to minimise the computational cost, whilst remaining within the target acceptance range.
When there are more step-sizes involved, the ratios between the remaining step-sizes are set relative to the second-coarsest scale via the force balancing method, such that we only have to tune the two coarsest scales.

\begin{figure}[tb]
\centering
\begin{tikzpicture}[trim axis group left]

\begin{groupplot}[
	mysmall,
	group style={
		group size=2 by 1,
		horizontal sep=5pt,
		yticklabels at=edge left,
		},
	cycle list name=mstone_d,
	ymin=0, ymax=80000,
	scaled y ticks=base 10:-4,
	ytick scale label code/.code={},
	minor y tick num=1,
	ymajorgrids,
	y errors,
	xlabel absolute group,
	]
	

\nextgroupplot[
	title=PF-RHMC,
	xlabel=$p$,
	ylabel=$C / 10^4$,
	only marks,
]

\renewcommand{\matrixopfile}{\figdir/data/PFRHMC/witers_j135X_cs.txt}

\addplot
	table[
		x=p,
		y=iter_SF,
		y error expr={\thisrow{iter_SF_err}*\autocorr},
	]
	{\matrixopfile};

\foreach \i in {1,2} {
\addplot
	table[
		x=p,
		y=iter_F\i,
		y error expr={\thisrow{iter_F\i_err}*\autocorr},
	]
	{\matrixopfile};
}

\filldraw[draw=band_1_out, fill=band_1]
 (2, 68060-\autocorr*1170) rectangle (18, 68060+\autocorr*1170);
\draw[band_1_cen_light] (2, 68060) -- (18, 68060);

\addplot+[red, mark=*]
	table[
		x=p,
		y=iter_tot,
		y error expr={\thisrow{iter_tot_err}*\autocorr},
	]
	{\matrixopfile};


\nextgroupplot[
	title=tRHMC,
	xlabel=$t$,
	only marks,
]

\renewcommand{\matrixopfile}{\figdir/data/tRHMC/witers_j138X_cs.txt}

\addplot
	table[
		x=t,
		y=iter_SF,
		y error expr={\thisrow{iter_SF_err}*\autocorr},
	]
	{\matrixopfile};

\foreach \i in {1,2} {
\addplot
	table[
		x=t,
		y=iter_F\i,
		y error expr={\thisrow{iter_F\i_err}*\autocorr},
	]
	{\matrixopfile};
}

\filldraw[draw=band_1_out, fill=band_1]
 (0, 68060-\autocorr*1170) rectangle (10, 68060+\autocorr*1170);
\draw[band_1_cen_light] (0, 68060) -- (10, 68060);

\addplot+[red, mark=*]
	table[
		x=t,
		y=iter_tot,
		y error expr={\thisrow{iter_tot_err}*\autocorr},
	]
	{\matrixopfile};

\end{groupplot}

\end{tikzpicture}
\caption{C-scale tuned cost for the single-filter actions on the $16^3 \times 32$ lattice.
Refer to Figure \ref{fig:1f_rhmc_cost} for the legend.
\label{fig:1f_cs_cost}}
\end{figure}
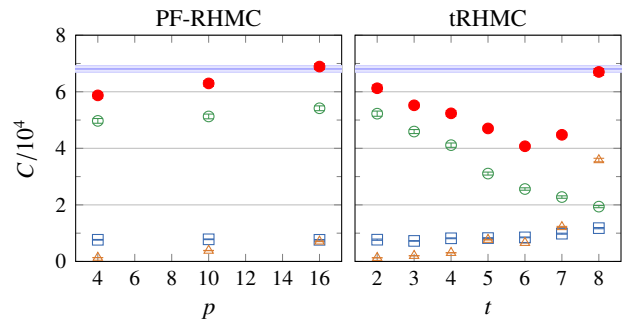

We performed this process for PF-RHMC and tRHMC, varying $n_1$ for each force-balanced configuration to find the minima in cost.
The resultant configurations are shown in Tables~\ref{tbl:pfrhmc_cs} and~\ref{tbl:trhmc_cs}  and the cost function is plotted in Figure~\ref{fig:1f_cs_cost}.

The cost for PF-RHMC has no significant improvement under c-scale tuning (compare with Figure \ref{fig:1f_rhmc_cost}).
We note that the step-sizes under c-scale tuning are not very different from those of force balancing (see Tables \ref{tbl:pfrhmc} and \ref{tbl:pfrhmc_cs}).
This indicates that force balancing gets us close to the optimal parameter set for PF-RHMC, which is perhaps not so surprising given that the polynomial filter does not drastically decrease the maximal force $F_2$ due to the correction term.

The effect of c-scale tuning on tRHMC is much more pronounced, particularly at higher truncation orders where the maximal force due to the correction term becomes very small. While we do not get much of an improvement in cost for the previous best tRHMC point $t=4$, we do engineer a significant reduction in the cost for $t=2,5,6,8$. The new optimal point is $t=6$ with cost $C = 40,700 \pm 700$, now $30\%$ cheaper than plain RHMC.

Note that under c-scale tuning the tRHMC computational cost for $t=4,5,6$  is similar, and the cost for $t=8$ is vastly improved over force balancing.
This demonstrates that characteristic scale tuning works, in that it reduces filter parameter sensitivity. In practice, this means that using c-scale tuning should reduce the need for re-tuning the filter parameter(s) when generating a new lattice with different physical parameters (such as the quark mass).

Up to this point, we have considering adding a single filter term for both PF-RHMC and tRHMC.
When more than one filter is applied, the filter parameter space is larger, and the range of force magnitudes that needs to be considered also increases. 
However, using the characteristic scale tuning technique above, we still only need to tune the two coarsest step-sizes, the remainder are set by force-balancing.
As we will demonstrate in the next section, when applying multiple filters the reduced sensitivity to the filter parameters has a significant benefit; one can choose a reasonable set of filters, then apply c-scale tuning to get near-optimal computational cost.

\subsection{Two-filter results}

The use of multiple hierarchical filters has been shown to be beneficial for degenerate two-flavour simulations in previous studies, see e.g.~\cite{Kamleh:2011dc,Haar:2016bwe,Aoki:2009ix,Bruno:2014jqa}.
We now consider using multiple filters for RHMC: to test if they will similarly improve the computational performance, and to measure the effectiveness of the c-scale tuning proposed above.

We first consider polynomial filtering. For two polynomial filters (2PF-RHMC), we select a pair of polynomials $P(K), Q(K)$ such that $P(K) \approx K^{-1/2}$ and $Q(K) \approx P(K)^{-1} K^{-1/2}$. This yields the action
\begin{equation}
	S_{2PF-RHMC} = \phi_1^\dag P(K) \phi_1 + \phi_2^\dag Q(K) \phi_2 + \phi_3^\dag [P(K) Q(K)]^{-1}R(K) \phi_3 \label{eq:2pfrhmc}.
\end{equation}
The 2PF-RHMC fermion action has two polynomial orders $p, q$ and three step-sizes $\{h_1, h_2, h_3\}$ to tune. We fix $p = 4$, vary $q$, and then tune the step-sizes with force balancing.
This gives the configurations in Table~\ref{tbl:2pfrhmc_fdt}.
Starting from this parameter set, we perform c-scale tuning by varying the second-coarsest step-size $h_2$ while keeping $h_2/h_1$ fixed.
The resultant configuration choices are given in Table~\ref{tbl:2pfrhmc}.

For two-level truncation filtering (2tRHMC), we use a pair of truncation orders $t < t'$ giving fermion action
\begin{equation}
	S_{2tRHMC} = \phi_1^\dag R_{0,t}(K) \phi_1 + \phi_2^\dag R_{t,t'}(K) \phi_2 + \phi_3^\dag R_{t',n}(K) \phi_3. \label{eq:2trhmc}
\end{equation}

We set $t = \{4,5\}$, vary $t' > t$, and apply force balancing and c-scale tuning to the step-sizes. The resultant configuration choices are given in Table \ref{tbl:2trhmc_fdt} (force balancing) and Table \ref{tbl:2trhmc} (c-scale tuning).

\begin{figure}[tb]
\centering
\begin{tikzpicture}[trim axis group left]

\begin{groupplot}[
	mysmall,
	group style={
		group size=3 by 1,
		horizontal sep=5pt,
		yticklabels at=edge left,
		},
	cycle list name=mstone_d,
	ymin=0, ymax=140000,
	scaled y ticks=base 10:-4,
	ytick scale label code/.code={},
	minor y tick num=1,
	ymajorgrids,
	y errors,
	xlabel absolute group,
	ylabel shift=-0.5em,
	]

\nextgroupplot[
	title={2PF-RHMC\makebox[0pt]{\phantom{$t_1$}}},
	xlabel={$q$\makebox[0pt]{\phantom{$t$}}},
	ylabel={$C / 10^4$},
	only marks,
	width=2.5cm,
]

\renewcommand{\matrixopfile}{\figdir/data/2PFRHMC/witers_j136X.txt}

\filldraw[draw=band_1_out, fill=band_1]
 (8, 68060-\autocorr*1170) rectangle (22, 68060+\autocorr*1170);
\draw[band_1_cen_light] (8, 68060) -- (22, 68060);

\addplot
	table[
		x=q,
		y=iter_SF,
		y error expr={\thisrow{iter_SF_err}*\autocorr},
	]
	{\matrixopfile};

\foreach \i in {1,2,3} {
\addplot
	table[
		x=q,
		y=iter_F\i,
		y error expr={\thisrow{iter_F\i_err}*\autocorr},
	]
	{\matrixopfile};
}

\addplot+[red, mark=*]
	table[
		x=q,
		y=iter_tot,
		y error expr={\thisrow{iter_tot_err}*\autocorr},
	]
	{\matrixopfile};


\nextgroupplot[
	title={2tRHMC, $t = 4$},
	xlabel={$t'$},
	only marks,
	xtick={5,6,7,8},
	width=2.5cm,
]

\renewcommand{\matrixopfile}{\figdir/data/2tRHMC/witers_j1380X_t4.txt}

\filldraw[draw=band_1_out, fill=band_1]
 (4, 68060-\autocorr*1170) rectangle (9, 68060+\autocorr*1170);
\draw[band_1_cen_light] (4, 68060) -- (9, 68060);

\addplot
	table[
		x=t2,
		y=iter_SF,
		y error expr={\thisrow{iter_SF_err}*\autocorr},
	]
	{\matrixopfile};

\foreach \i in {1,2,3} {
\addplot
	table[
		x=t2,
		y=iter_F\i,
		y error expr={\thisrow{iter_F\i_err}*\autocorr},
	]
	{\matrixopfile};
}

\addplot+[red, mark=*]
	table[
		x=t2,
		y=iter_tot,
		y error expr={\thisrow{iter_tot_err}*\autocorr},
	]
	{\matrixopfile};


\nextgroupplot[
	title={2tRHMC, $t=5$},
	xlabel={$t'$},
	only marks,
	xtick={7,8,9,10},
	width=2.5cm,
]

\renewcommand{\matrixopfile}{\figdir/data/2tRHMC/witers_j1380X_t5.txt}

\filldraw[draw=band_1_out, fill=band_1]
 (6, 68060-\autocorr*1170) rectangle (11, 68060+\autocorr*1170);
\draw[band_1_cen_light] (6, 68060) -- (11, 68060);

\addplot
	table[
		x=t2,
		y=iter_SF,
		y error expr={\thisrow{iter_SF_err}*\autocorr},
	]
	{\matrixopfile};

\foreach \i in {1,2,3} {
\addplot
	table[
		x=t2,
		y=iter_F\i,
		y error expr={\thisrow{iter_F\i_err}*\autocorr},
	]
	{\matrixopfile};
}

\addplot+[red, mark=*]
	table[
		x=t2,
		y=iter_tot,
		y error expr={\thisrow{iter_tot_err}*\autocorr},
	]
	{\matrixopfile};


\end{groupplot}

\end{tikzpicture}
\caption{The cost function for 2PF-RHMC and 2tRHMC on the $16^3 \times 32$ lattice using force balancing.
Filled circles are the total cost.
Empty squares are the component of the total cost due to action initialisation,
empty triangles due to calculating $F_1$,
empty circles due to calculating $F_2$,
and empty diamonds due to calculating $F_3$.
The faded band is the cost of plain RHMC, included for ease of comparison.
\label{fig:2f_cost_fb}}
\end{figure}
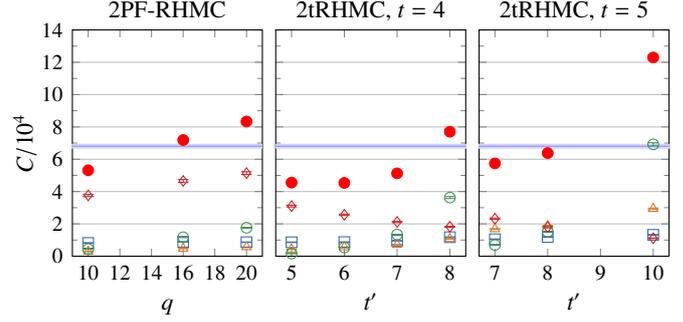

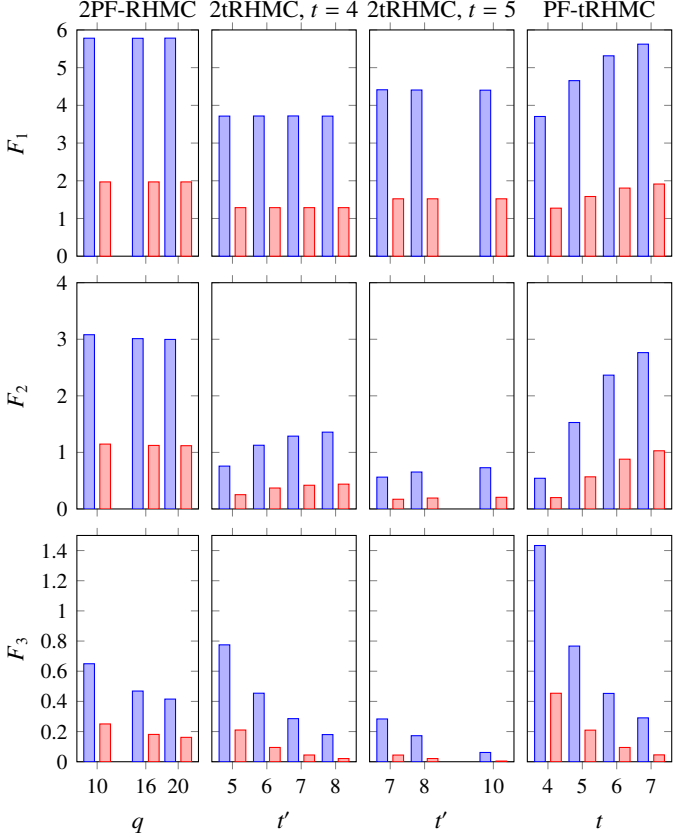
\begin{figure}[tb]
\centering
\begin{tikzpicture}[baseline, trim axis group left]

\renewcommand{\maxforcefile}{\figdir/data/2PFRHMC/fmax_j136X.txt}
\renewcommand{\avgforcefile}{\figdir/data/2PFRHMC/favg_j136X.txt}
\newcommand{\maxforcefilea}{\figdir/data/2tRHMC/fmax_j1380X_t4.txt}
\newcommand{\avgforcefilea}{\figdir/data/2tRHMC/favg_j1380X_t4.txt}
\newcommand{\maxforcefileb}{\figdir/data/PFtRHMC/fmax_j1350X.txt}
\newcommand{\avgforcefileb}{\figdir/data/PFtRHMC/favg_j1350X.txt}
\newcommand{\maxforcefilec}{\figdir/data/2tRHMC/fmax_j1380X_t5.txt}
\newcommand{\avgforcefilec}{\figdir/data/2tRHMC/favg_j1380X_t5.txt}

\pgfplotsset{
	left plot/.style={
		bar width=4pt,
		xtick=data,
		enlarge x limits=0.25,
		width=1.6cm,
	},
	centre plot/.style={
		bar width=4pt,
		xtick=data,
		enlarge x limits=0.2,
		width=1.9cm,
	},
	right plot/.style={
		bar width=4pt,
		xtick=data,
		enlarge x limits=0.2,
		width=1.9cm,
	},
}

\begin{groupplot}[
	mysmall,
	group style={
		group size=4 by 3,
		xlabels at=edge bottom,
		horizontal sep=5pt,
		vertical sep=10pt,
		xticklabels at=edge bottom,
		yticklabels at=edge left,
		},
	ybar,
	xlabel absolute group=-2.5em,
	ylabel absolute group,
	]


\nextgroupplot[
	title={2PF-RHMC\makebox[0pt]{\phantom{,}}},
	ylabel={$F_1$},
	ymin=0, ymax=6,
	left plot,
]

\addplot
	table[
		x=q,
		y=F_F1,
	]
	{\maxforcefile};

\addplot
	table[
		x=q,
		y=F_F1,
	]
	{\avgforcefile};

\nextgroupplot[
	title={2tRHMC, $t=4$},
	ymin=0, ymax=6,
	centre plot,
]

\addplot
	table[
		x=t2,
		y=F_F1,
	]
	{\maxforcefilea};

\addplot
	table[
		x=t2,
		y=F_F1,
	]
	{\avgforcefilea};

\nextgroupplot[
	title={2tRHMC, $t=5$},
	ymin=0, ymax=6,
	centre plot,
]

\addplot
	table[
		x=t2,
		y=F_F1,
	]
	{\maxforcefilec};

\addplot
	table[
		x=t2,
		y=F_F1,
	]
	{\avgforcefilec};

\nextgroupplot[
	title={PF-tRHMC\makebox[0pt]{\phantom{,}}},
	ymin=0, ymax=6,
	right plot,
]

\addplot
	table[
		x=t,
		y=F_F1,
	]
	{\maxforcefileb};

\addplot
	table[
		x=t,
		y=F_F1,
	]
	{\avgforcefileb};


\nextgroupplot[
	ylabel={$F_2$},
	ymin=0, ymax=4,
	left plot,
]

\addplot
	table[
		x=q,
		y=F_F2,
	]
	{\maxforcefile};

\addplot
	table[
		x=q,
		y=F_F2,
	]
	{\avgforcefile};

\nextgroupplot[
	ymin=0, ymax=4,
	centre plot,
]

\addplot
	table[
		x=t2,
		y=F_F2,
	]
	{\maxforcefilea};

\addplot
	table[
		x=t2,
		y=F_F2,
	]
	{\avgforcefilea};

\nextgroupplot[
	ymin=0, ymax=4,
	centre plot,
]

\addplot
	table[
		x=t2,
		y=F_F2,
	]
	{\maxforcefilec};

\addplot
	table[
		x=t2,
		y=F_F2,
	]
	{\avgforcefilec};

\nextgroupplot[
	ymin=0, ymax=4,
	right plot,
]

\addplot
	table[
		x=t,
		y=F_F2,
	]
	{\maxforcefileb};

\addplot
	table[
		x=t,
		y=F_F2,
	]
	{\avgforcefileb};


\nextgroupplot[
	ylabel={$F_3$},
	xlabel={$q$},
	ymin=0, ymax=1.5,
	left plot,
]

\addplot
	table[
		x=q,
		y=F_F3,
	]
	{\maxforcefile};

\addplot
	table[
		x=q,
		y=F_F3,
	]
	{\avgforcefile};

\nextgroupplot[
	xlabel={$t'$},
	ymin=0, ymax=1.5,
	centre plot,
]

\addplot
	table[
		x=t2,
		y=F_F3,
	]
	{\maxforcefilea};

\addplot
	table[
		x=t2,
		y=F_F3,
	]
	{\avgforcefilea};

\nextgroupplot[
	xlabel={$t'$},
	ymin=0, ymax=1.5,
	centre plot,
]

\addplot
	table[
		x=t2,
		y=F_F3,
	]
	{\maxforcefilec};

\addplot
	table[
		x=t2,
		y=F_F3,
	]
	{\avgforcefilec};

\nextgroupplot[
	xlabel={$t$},
	ymin=0, ymax=1.5,
	right plot,
]

\addplot
	table[
		x=t,
		y=F_F3,
	]
	{\maxforcefileb};

\addplot
	table[
		x=t,
		y=F_F3,
	]
	{\avgforcefileb};

\end{groupplot}
\end{tikzpicture}
\caption{Forces for the 2-filter actions on the $16^3 \times 32$ lattice.
Left-hand bars show the maximal force, while right-hand bars show the average. \label{fig:force_2f}}
\end{figure}

The cost function for 2PF-RHMC and 2tRHMC using force balancing is shown in Figure \ref{fig:2f_cost_fb}, and the corresponding forces are shown in Figure \ref{fig:force_2f}.
Comparing with the one-filter case Figure \ref{fig:1f_rhmc_cost},
we only see a small additional benefit to using two polynomial filters versus just one, with a minima at $p=4, q=10$.
Looking at the tRHMC results, using two truncations with $t=4$ keeps the cost at a level similar to the best 1tRHMC point for $t' = 5,6,7.$
However, as in the single filter case, using force-balancing to set the step sizes is far from optimal for higher polynomial orders, see $q=16,20$, or at larger truncation orders, see $(t,t') = (4,8)$ or any of $t=5$ actions.

\begin{figure}[tbp]
\centering
\begin{tikzpicture}[trim axis group left]

\begin{groupplot}[
	mysmall,
	group style={
		group size=3 by 1,
		horizontal sep=5pt,
		yticklabels at=edge left,
		},
	cycle list name=mstone_d,
	ymin=0, ymax=80000,
	scaled y ticks=base 10:-4,
	ytick scale label code/.code={},
	minor y tick num=1,
	ymajorgrids,
	y errors,
	xlabel absolute group,
	width=2.5cm,
	]

\nextgroupplot[
	title={2PF-RHMC\makebox[0pt]{\phantom{$t_1$}}},
	xlabel={$q$\makebox[0pt]{\phantom{$t$}}},
	ylabel={$C / 10^4$},
	only marks,
]

\renewcommand{\matrixopfile}{\figdir/data/2PFRHMC/witers_j136X_cs.txt}

\filldraw[draw=band_1_out, fill=band_1]
 (8, 68060-\autocorr*1170) rectangle (22, 68060+\autocorr*1170);
\draw[band_1_cen_light] (8, 68060) -- (22, 68060);

\addplot
	table[
		x=q,
		y=iter_SF,
		y error expr={\thisrow{iter_SF_err}*\autocorr},
	]
	{\matrixopfile};

\foreach \i in {1,2,3} {
\addplot
	table[
		x=q,
		y=iter_F\i,
		y error expr={\thisrow{iter_F\i_err}*\autocorr},
	]
	{\matrixopfile};
}

\addplot+[red, mark=*]
	table[
		x=q,
		y=iter_tot,
		y error expr={\thisrow{iter_tot_err}*\autocorr},
	]
	{\matrixopfile};


\nextgroupplot[
	title={2tRHMC, $t=4$},
	xlabel={$t'$},
	only marks,
	xtick={5,6,7,8},
]

\renewcommand{\matrixopfile}{\figdir/data/2tRHMC/witers_j1380X_t4_cs.txt}

\filldraw[draw=band_1_out, fill=band_1]
 (4, 68060-\autocorr*1170) rectangle (9, 68060+\autocorr*1170);
\draw[band_1_cen_light] (4, 68060) -- (9, 68060);

\addplot
	table[
		x=t2,
		y=iter_SF,
		y error expr={\thisrow{iter_SF_err}*\autocorr},
	]
	{\matrixopfile};

\foreach \i in {1,2,3} {
\addplot
	table[
		x=t2,
		y=iter_F\i,
		y error expr={\thisrow{iter_F\i_err}*\autocorr},
	]
	{\matrixopfile};
}

\addplot+[red, mark=*]
	table[
		x=t2,
		y=iter_tot,
		y error expr={\thisrow{iter_tot_err}*\autocorr},
	]
	{\matrixopfile};


\nextgroupplot[
	title={2tRHMC, $t=5$},
	xlabel={$t'$},
	only marks,
	xtick={7,8,9,10},
]

\renewcommand{\matrixopfile}{\figdir/data/2tRHMC/witers_j1380X_t5_cs.txt}

\filldraw[draw=band_1_out, fill=band_1]
 (6, 68060-\autocorr*1170) rectangle (11, 68060+\autocorr*1170);
\draw[band_1_cen_light] (6, 68060) -- (11, 68060);

\addplot
	table[
		x=t2,
		y=iter_SF,
		y error expr={\thisrow{iter_SF_err}*\autocorr},
	]
	{\matrixopfile};

\foreach \i in {1,2,3} {
\addplot
	table[
		x=t2,
		y=iter_F\i,
		y error expr={\thisrow{iter_F\i_err}*\autocorr},
	]
	{\matrixopfile};
}

\addplot+[red, mark=*]
	table[
		x=t2,
		y=iter_tot,
		y error expr={\thisrow{iter_tot_err}*\autocorr},
	]
	{\matrixopfile};


\end{groupplot}

\end{tikzpicture}
\caption{C-scale tuned cost for 2PF-RHMC and 2tRHMC on the $16^3 \times 32$ lattice.
Refer to Figure \ref{fig:2f_cost_fb} for the legend.
\label{fig:2f_cost}}
\end{figure}
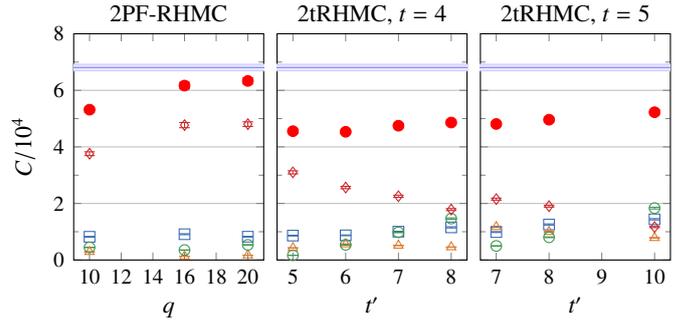

Next we consider c-scale tuning for the two-filter actions, with cost functions shown in Figure~\ref{fig:2f_cost}.
Comparing this with the force balanced case (Figure~\ref{fig:2f_cost_fb}), 2PF-RHMC gets some improvement in the case of higher order polynomials $q=16,20$, but overall there is only a relatively small benefit.
The optimal point $(p,q) = (4,10)$ has a cost $C = 53,200 \pm 900$, yielding a $22\%$ improvement over plain RHMC.

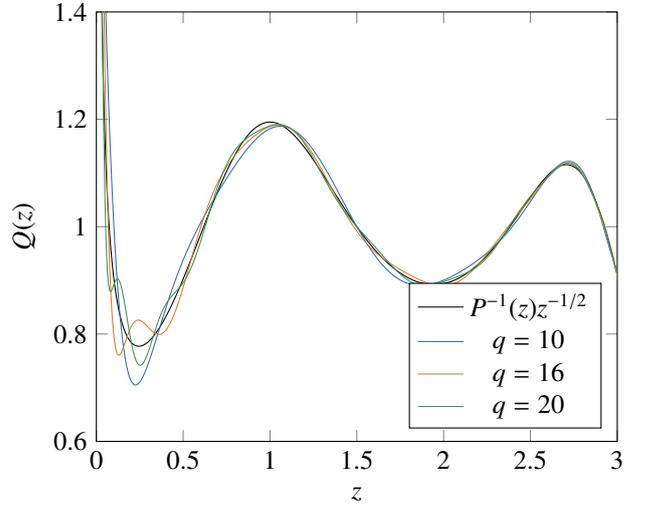
\begin{figure}[tbp]
\centering
\begin{tikzpicture}[trim axis left]

\begin{axis}[
	cycle list name=mstone_d,
	xmin=0, xmax=3,
	ymin=0.6, ymax=1.4,
	restrict y to domain*=0.5:1.5,
	xlabel=$z$,
	ylabel=$Q(z)$,
	legend pos=south east,
	]

\renewcommand{\matrixopfile}{\figdir/data/2PFRHMC/2pfrhmc_filter.txt}

\addplot+[black,mark=none]
	table[
		x=x,
		y=p4r,
	]
	{\matrixopfile};
	
\pgfplotsset{cycle list shift=-1}

\foreach \i in {10,16,20} {
\addplot+[mark=none]
	table[
		x=x,
		y=q\i,
	]
	{\matrixopfile};
}

\legend{$P^{-1}(z) z^{-1/2}$, $q=10$, $q=16$, $q=20$}

	]

\end{axis}

\end{tikzpicture}
\caption{$Q(z)$ for various intermediate polynomial orders $q$ and $p=4$, comparing them to that which they approximate, $P^{-1}(z) z^{-1/2}$ with $p=4$. \label{fig:2pfrhmc_filter}}
\end{figure}

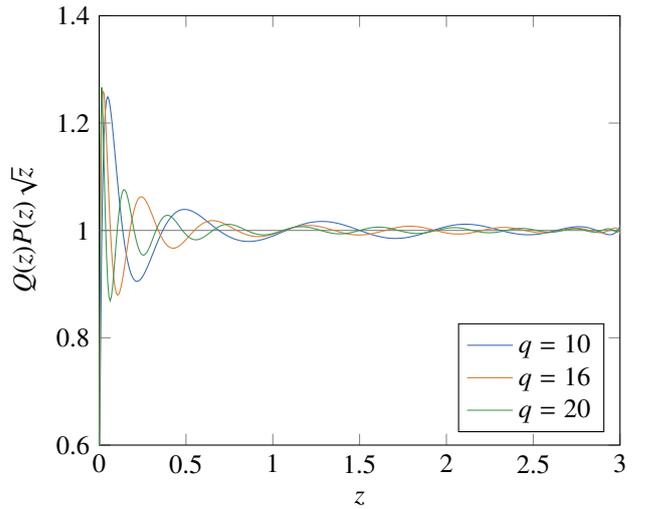
\begin{figure}[tbp]
\centering
\begin{tikzpicture}[trim axis left]

\begin{axis}[
	xmin=0, xmax=3,
	ymin=0.6, ymax=1.4,
	restrict y to domain*=0.5:1.5,
	cycle list name=mstone_d,
	xlabel=$z$,
	ylabel=$Q(z)P(z)\sqrt{z}$,
	legend pos=south east,
	]

\renewcommand{\matrixopfile}{\figdir/data/2PFRHMC/2pfrhmc_corr.txt}

\draw[thin, gray]
	(0,1) -- (3,1);

\foreach \i in {10,16,20} {
\addplot+[mark=none]
	table[
		x=x,
		y=q\i,
	]
	{\matrixopfile};
}

\legend{$q=10$, $q=16$, $q=20$}

\end{axis}

\end{tikzpicture}
\caption{$Q(z)P(z)\sqrt{z}$ for intermediate polynomial orders $q$ and $p=4$.
$R(z)$ is the 20th order Zolotarev approximation given in Table \ref{tbl:zolo_eg}.
Closeness to unity indicates how well the intermediate filter $Q(K)$ approximates $P^{-1}(K) R(K)$. \label{fig:2pfrhmc_correction}}
\end{figure}

The main contributor to the 2PF-RHMC cost is, as in the single-filter case, the cost of evaluating the correction term $F_3$ (empty diamonds in Figure \ref{fig:2f_cost}), which is symptomatic of an increasing step count $n_3$ (see Table~\ref{tbl:2pfrhmc}).
Looking at $Q(K)$ in Figure~\ref{fig:2pfrhmc_filter} and the relative error of $Q(K)$ in approximating $P^{-1}(K) R(K)$ in Figure~\ref{fig:2pfrhmc_correction}, we find that increasing the polynomial order $q$ does not greatly improve the approximation, especially at small $K$.
This indicates that $S_2$ does not capture more of the action as we increase $q$, leaving the brunt of the dynamics and required computational work at the expensive correction term $S_3$.
This can explain why PF-RHMC performs poorly as well, for the error in the approximation $P(K)$ is large at small $K$ (see Figure \ref{fig:pfrhmc_correction}).
This behaviour shows that using a Chebyshev polynomial approximation for PF-RHMC is ineffective.
While a different class of polynomials could have better approximation behaviour at lower values, the presence of the simpler tRHMC algorithm devalued an investigation at this time.

The optimal point for 2tRHMC with c-scale tuning is at $(t,t')=(4,6)$ with cost $C = 45,300 \pm 700$, a $32\%$ improvement over plain RHMC.
As was the case for a single truncation filter, the optimal cost is only marginally improved over the force balanced results, but the cost function under c-scale tuning is now consistent over all the filter choices shown.
This behaviour demonstrates that c-scale tuning reduces the sensitivity of the cost to the truncation order in the case of multiple tRHMC filters.

It is also possible to combine the two filtering techniques we consider here, applying both a polynomial filter and a truncation filter. For our PF-tRHMC tests, we place a $p=4$ polynomial filter on top of the tRHMC action, 
\begin{equation}
	S_{PF-tRHMC} = \phi_1^\dag P(K) \phi_1 + \phi_2^\dag P(K)^{-1} R_{0,t}(K) \phi_2 + \phi_3^\dag R_{t,n}(K) \phi_3 \label{eq:pftrhmc}.
\end{equation}

\begin{figure}[tbp]
\centering
\begin{tikzpicture}[trim axis group left, trim axis group right]

\begin{groupplot}[
	mysmall,
	group style={
		group size=2 by 1,
		horizontal sep=5pt,
		yticklabels at=edge left,
		},
	y errors,
	ymajorgrids,
	xtick=data,
	cycle list name=mstone_d,
	minor y tick num=1,
	scaled y ticks=base 10:-4,
	ytick scale label code/.code={},
]

\nextgroupplot[
	title=PF-tRHMC (F),
	xlabel=$t$,
	ylabel=$C / 10^4$,
	only marks,
	ymin=0, ymax=80000,
]

\renewcommand{\matrixopfile}{\figdir/data/PFtRHMC/witers_j1350X.txt}

\filldraw[draw=band_1_out, fill=band_1]
 (3, 68060-\autocorr*1170) rectangle (8, 68060+\autocorr*1170);
\draw[band_1_cen_light] (3, 68060) -- (8, 68060);

\addplot
	table[
		x=t,
		y=iter_SF,
		y error expr={\thisrow{iter_SF_err}*\autocorr},
	]
	{\matrixopfile};

\foreach \i in {1,2,3} {
\addplot
	table[
		x=t,
		y=iter_F\i,
		y error expr={\thisrow{iter_F\i_err}*\autocorr},
	]
	{\matrixopfile};
}

\addplot+[red, mark=*]
	table[
		x=t,
		y=iter_tot,
		y error expr={\thisrow{iter_tot_err}*\autocorr},
	]
	{\matrixopfile};



\nextgroupplot[
	title=PF-tRHMC (cs),
	xlabel=$t$,
	only marks,
	ymin=0, ymax=80000,
]

\renewcommand{\matrixopfile}{\figdir/data/PFtRHMC/witers_j1350X_cs.txt}

\filldraw[draw=band_1_out, fill=band_1]
 (3, 68060-\autocorr*1170) rectangle (8, 68060+\autocorr*1170);
\draw[band_1_cen_light] (3, 68060) -- (8, 68060);

\addplot
	table[
		x=t,
		y=iter_SF,
		y error expr={\thisrow{iter_SF_err}*\autocorr},
	]
	{\matrixopfile};

\foreach \i in {1,2,3} {
\addplot
	table[
		x=t,
		y=iter_F\i,
		y error expr={\thisrow{iter_F\i_err}*\autocorr},
	]
	{\matrixopfile};
}

\addplot+[red, mark=*]
	table[
		x=t,
		y=iter_tot,
		y error expr={\thisrow{iter_tot_err}*\autocorr},
	]
	{\matrixopfile};

\end{groupplot}

\end{tikzpicture}
\caption{The cost function for PF-tRHMC on the $16^3 \times 32$ lattice.
The left hand plot shows the force-balanced data, while the right hand plot shows the c-scale tuned data.
Refer to Figure \ref{fig:2f_cost_fb} for the legend.
\label{fig:pftrhmc_cost}}
\end{figure}
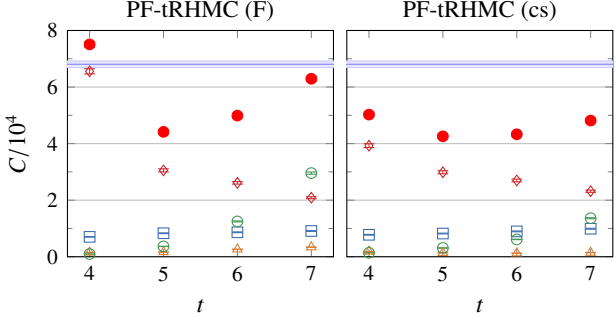

This form suggests using a polynomial $P(K)$ that approximates $R_{0,t}(K)^{-1}$. As before, we make use of a Chebyshev approximation.
The configuration choices are shown in Tables~\ref{tbl:pftrhmc_fdt} and \ref{tbl:pftrhmc} for force balancing and c-scale tuning respectively, the corresponding forces on the right-hand side in Figure \ref{fig:force_2f}, and the cost function in Figure \ref{fig:pftrhmc_cost}.

PF-tRHMC with force balancing is very sensitive to the choice of truncation order $t$, with inflated cost at $t=4$ and $t=7$.
This is appears to be due to a rapidly varying distribution of forces between the polynomial and truncated terms in the action (see Figure \ref{fig:force_2f}).

When we apply c-scale tuning, PF-tRHMC performs as well as (or slightly better than) 2tRHMC, with a relatively consistent cost improvement over the range of filters tested $t=4,5,6,7$.
The optimal filter choice is $(p,t) = (4,5)$ with cost $C = 42,600 \pm 700$, which is a small improvement on the optimal 2tRHMC result.
The superior behaviour of PF-tRHMC, compared with 2PF-RHMC, is due to the improved performance of the polynomial filter correction term (empty circles in Figure \ref{fig:pftrhmc_cost}).
This occurs because the polynomial filter is small ($p=4$) and does not have to approximate the full inverse square-root.
However, we note that 2tRHMC provides a similar benefit whilst being simpler to implement because it only has one type of filtering.

\subsection{Larger lattice tests}

In order to determine whether the single-flavour improvement techniques described here scale to more physical lattices, we also run comparison tests on a $24^3 \times 48$ lattice with two degenerate single-flavour clover-improved fermions, pion mass $m_\pi \approx 300\ \mathrm{MeV}$ and lattice spacing $\sim 0.07$~fm~\cite{Bali:2012qs}.
Based on the relative merits of the  $16^3 \times 32$ results, we do not consider polynomial filtering on the larger volume, and only compare plain RHMC, tRHMC and 2tRHMC.
The cost of simulating the $24^3 \times 48$ lattice is much larger than for the $16^2 \times 32$ lattice, so we only have $\sim 100$ trajectories per filter set.
Nonetheless, these statistics are sufficient to compare the different filtering methods to the baseline result.

For the rational approximation on this lattice, we use the 30th order Zolotarev approximation optimised the range $[10^{-6}, 3]$.
When we measure the eigenvalue spectrum of the fermion matrix, we find that the distribution extends slightly outside this range; $[\lambda_{\mathrm{min}}, \lambda_{\mathrm{max}}] = [3.3(1) \times 10^{-6}, 3.078(4)]$.
Nonetheless, the rational function is still valid as the error of the Zolotarev approximation remains within the desired tolerance well above the upper bound for this range.

We note that using clover-improved fermions under even-odd preconditioning adds an extra term to the fermion action, the determinant term $S_{\mathrm{det}} = - 2 \tr(\ln(D_{ee}))$, which much be placed on an integration scale.
This term is relatively cheap to calculate, so for this work it is always integrated on the second-finest integration scale, i.e. $n_1$.

Using $n_1 = 35$ steps for the pseudofermion and determinant terms, and $n_0 = 480$ for the gauge term, plain RHMC for the $24^3 \times 48$ lattice takes $N_{\mathrm{mat}} = 628,000 \pm 4,000$ matrix operations per trajectory with an acceptance rate of $P_{acc} = 0.65(4)$. This gives a baseline cost of
\begin{equation}
	C = 971,000 \pm 65,000
\end{equation}
for comparison.

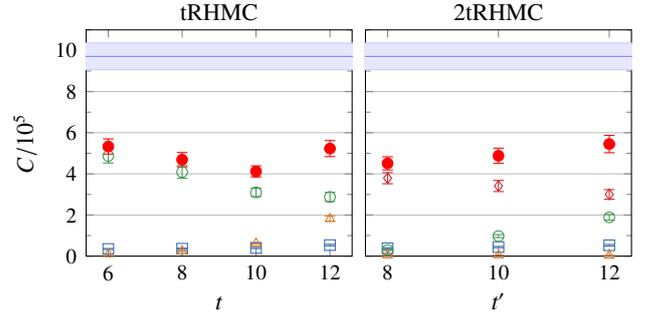
\begin{figure}[tbp]
\centering
\begin{tikzpicture}[trim axis group left]

\begin{groupplot}[
	mysmall,
	group style={
		group size=2 by 1,
		horizontal sep=5pt,
		yticklabels at=edge left,
		},
	cycle list name=mstone_d,
	ymin=0, ymax=1100000,
	scaled y ticks=base 10:-5,
	ytick scale label code/.code={},
	minor y tick num=1,
	ymajorgrids,
	y errors,
	xlabel absolute group,
	]

\nextgroupplot[
	title={tRHMC},
	xlabel={$t$},
	ylabel={$C / 10^5$},
	only marks,
	xtick={4,6,8,10,12},
]

\renewcommand{\matrixopfile}{\figdir/data/24x48/tRHMC/witers_j1518X_cs.txt}

\filldraw[draw=band_1_out, fill=band_1]
 (-1, 971000-\autocorr*65000) rectangle (13, 971000+\autocorr*65000);
\draw[band_1_cen_light] (-1, 971000) -- (13, 971000);

\addplot
	table[
		x=t,
		y=iter_SF,
		y error expr={\thisrow{iter_SF_err}*\autocorr},
	]
	{\matrixopfile};

\foreach \i in {1,2} {
\addplot
	table[
		x=t,
		y=iter_F\i,
		y error expr={\thisrow{iter_F\i_err}*\autocorr},
	]
	{\matrixopfile};
}

\addplot+[red, mark=*]
	table[
		x=t,
		y=iter_tot,
		y error expr={\thisrow{iter_tot_err}*\autocorr},
	]
	{\matrixopfile};


\nextgroupplot[
	title={2tRHMC},
	xlabel={$t'$},
	only marks,
	xtick={6,8,10,12},
]

\renewcommand{\matrixopfile}{\figdir/data/24x48/2tRHMC/witers_j1519X_cs.txt}

\filldraw[draw=band_1_out, fill=band_1]
 (5, 971000-\autocorr*65000) rectangle (13, 971000+\autocorr*65000);
\draw[band_1_cen_light] (5, 971000) -- (13, 971000);

\addplot
	table[
		x=t2,
		y=iter_SF,
		y error expr={\thisrow{iter_SF_err}*\autocorr},
	]
	{\matrixopfile};

\foreach \i in {1,2,3} {
\addplot
	table[
		x=t2,
		y=iter_F\i,
		y error expr={\thisrow{iter_F\i_err}*\autocorr},
	]
	{\matrixopfile};
}

\addplot+[red, mark=*]
	table[
		x=t2,
		y=iter_tot,
		y error expr={\thisrow{iter_tot_err}*\autocorr},
	]
	{\matrixopfile};


\end{groupplot}

\end{tikzpicture}
\caption{C-scale tuned cost of tRHMC and 2tRHMC on the $24^3 \times 48$ lattice.
Filled circles are the total cost.
Empty squares are the component of the total cost due to action initialisation,
empty triangles due to calculating $F_1$,
empty circles due to calculating $F_2$,
and empty diamonds due to calculating $F_3$.
The faded band is the cost of plain RHMC, included for ease of comparison.
\label{fig:24x48_trhmc_cs_cost}}
\end{figure}

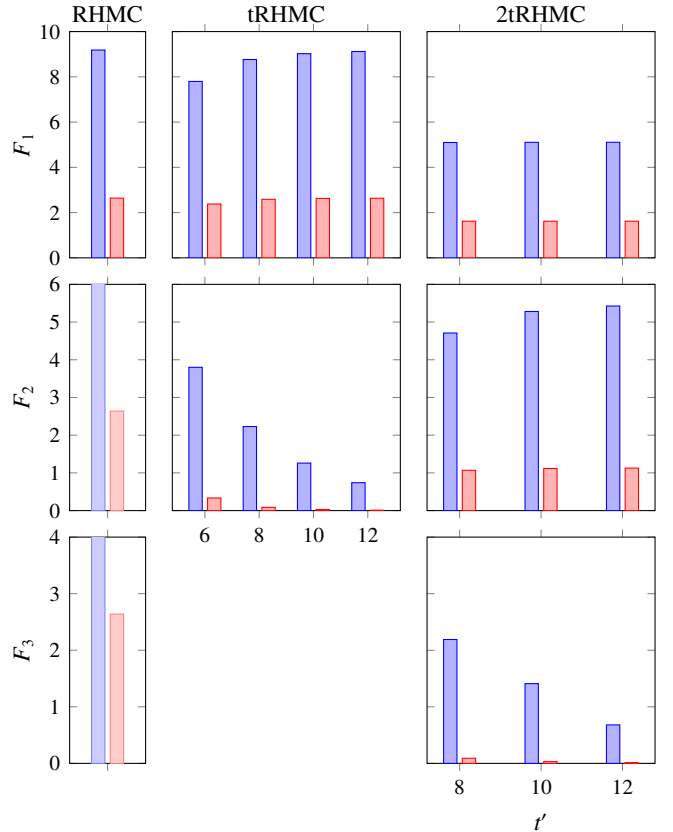
\begin{figure}[tbp]
\centering
\begin{tikzpicture}[baseline, trim axis group left]

\renewcommand{\maxforcefile}{\figdir/data/24x48/tRHMC/fmax_j1518X.txt}
\renewcommand{\avgforcefile}{\figdir/data/24x48/tRHMC/favg_j1518X.txt}
\newcommand{\maxforcefilea}{\figdir/data/24x48/2tRHMC/fmax_j1519X.txt}
\newcommand{\avgforcefilea}{\figdir/data/24x48/2tRHMC/favg_j1519X.txt}
\newcommand{\maxforcefileb}{\figdir/data/24x48/RHMC/fmax_j15150.txt}
\newcommand{\avgforcefileb}{\figdir/data/24x48/RHMC/favg_j15150.txt}

\pgfplotsset{
	left plot/.style={
		bar width=5pt,
		xtick=data,
		enlarge x limits=0.2,
		width=3cm,
	},
	right plot/.style={
		bar width=5pt,
		xtick=data,
		enlarge x limits=0.2,
		width=3cm,
	},
}

\begin{groupplot}[
	mysmall,
	group style={
		group size=3 by 3,
		xlabels at=edge bottom,
		horizontal sep=10pt,
		vertical sep=10pt,
		xticklabels at=edge bottom,
		yticklabels at=edge left,
		},
	ybar,
	xlabel absolute group=-2.5em,
	ylabel absolute group=1.5em,
	]


\nextgroupplot[
	title={RHMC},
	ylabel={$F_1$},
	ymin=0, ymax=10,
	left plot,
	width=1cm,
]

\addplot
	table[
		x=ID,
		y=F_F1,
		y error=F_F1_err,
	]
	{\maxforcefileb};

\addplot
	table[
		x=ID,
		y=F_F1,
		y error=F_F1_err,
	]
	{\avgforcefileb};

\nextgroupplot[
	title={tRHMC},
	ymin=0, ymax=10,
	left plot,
]

\addplot
	table[
		x=t,
		y=F_F1,
		y error=F_F1_err,
	]
	{\maxforcefile};

\addplot
	table[
		x=t,
		y=F_F1,
		y error=F_F1_err,
	]
	{\avgforcefile};

\nextgroupplot[
	title={2tRHMC},
	ymin=0, ymax=10,
	right plot,
]

\addplot
	table[
		x=t2,
		y=F_F1,
		y error=F_F1_err,
	]
	{\maxforcefilea};

\addplot
	table[
		x=t2,
		y=F_F1,
		y error=F_F1_err,
	]
	{\avgforcefilea};


\nextgroupplot[
	ylabel={$F_2$},
	ymin=0, ymax=6,
	left plot,
	width=1cm,
]

\addplot+[
	fill=blue!20!white,
	draw=blue!50!white,
]
	table[
		x=ID,
		y=F_F1,
		y error=F_F1_err,
	]
	{\maxforcefileb};

\addplot+[
	fill=red!20!white,
	draw=red!50!white,
]
	table[
		x=ID,
		y=F_F1,
		y error=F_F1_err,
	]
	{\avgforcefileb};

\nextgroupplot[
	ymin=0, ymax=6,
	left plot,
	xticklabels={6,8,10,12},
]

\addplot
	table[
		x=t,
		y=F_F2,
		y error=F_F2_err,
	]
	{\maxforcefile};

\addplot
	table[
		x=t,
		y=F_F2,
		y error=F_F2_err,
	]
	{\avgforcefile};

\nextgroupplot[
	ymin=0, ymax=6,
	right plot,
]

\addplot
	table[
		x=t2,
		y=F_F2,
		y error=F_F2_err,
	]
	{\maxforcefilea};

\addplot
	table[
		x=t2,
		y=F_F2,
		y error=F_F2_err,
	]
	{\avgforcefilea};


\nextgroupplot[
	ylabel={$F_3$},
	ymin=0, ymax=4,
	left plot,
	width=1cm,
	xticklabels={},
]

\addplot+[
	fill=blue!20!white,
	draw=blue!50!white,
]
	table[
		x=ID,
		y=F_F1,
		y error=F_F1_err,
	]
	{\maxforcefileb};

\addplot+[
	fill=red!20!white,
	draw=red!50!white,
]
	table[
		x=ID,
		y=F_F1,
		y error=F_F1_err,
	]
	{\avgforcefileb};


\nextgroupplot[group/empty plot]

\nextgroupplot[
	xlabel={$t'$},
	ymin=0, ymax=4,
	right plot,
]

\addplot
	table[
		x=t2,
		y=F_F3,
		y error=F_F3_err,
	]
	{\maxforcefilea};

\addplot
	table[
		x=t2,
		y=F_F3,
		y error=F_F3_err,
	]
	{\avgforcefilea};

\end{groupplot}
\end{tikzpicture}
\caption{Forces for RHMC, tRHMC and 2tRHMC on the $24^3 \times 48$ lattice.
Left-hand bars show the maximal force, while right-hand bars show the average. 
The single force term for plain RHMC is included for comparison on all force terms.
\label{fig:forces_24x48}}
\end{figure}

The cost functions for tRHMC and 2tRHMC (with $t=4$) using c-scale tuning are shown in Figure \ref{fig:24x48_trhmc_cs_cost}.
The baseline RHMC point is shown as a faded band.
The corresponding forces are shown in Figure \ref{fig:forces_24x48}, and the configuration data in Tables \ref{tbl:24x48_trhmc_cs} and \ref{tbl:24x48_2trhmc_cs}.

We find that the results with c-scale tuning are qualitatively similar to the $16^3 \times 32$ lattice results.
Once again we see that tRHMC provides the best result, with the minimum cost at $t=10$ with $C = 412,000 \pm 27,000$,
an impressive $58\%$ improvement over plain RHMC.
As was the case for the $16^3 \times 32$ lattice, we also see that using two truncation filters does not improve the minimum cost -- the optimal cost for 2tRHMC with $t=4$ is at $t'=8$, with $C = 451,000  \pm 32,000$.
Most importantly, we see that the cost is relatively flat across all filter parameters shown, demonstrating again that c-scale tuning reduces the need for filter parameter tuning.

\section{Conclusion}
\tikzsetfigurename{figure_4.}

We have studied two types of filtering that can be applied to RHMC: polynomial filtering (PF-RHMC) and truncated ordered product filtering (tRHMC). Comparing tRHMC and PF-RHMC for single-flavour pseudofermion simulations on a $16^3 \times 32$ lattice, we find that tRHMC performs better overall, providing a $40\%$ improvement in cost over the plain RHMC case.

We tested both one and two levels of filtering, finding that adding a second filter did not improve the performance.
When testing tRHMC on a larger $24^3 \times 48$ lattice at a lighter quark mass we find a similar improvement: tRHMC reduces the cost significantly, but two truncations are not better than one.
A possible explanation for this is that, by simulating a single flavour only (rather than two as a single pseudofermion), there is already a significant reduction in the force variance~\cite{Clark:2006fx}, and hence there is less work for the filter to do. 

The superior benefits of tRHMC over PF-RHMC are possibly due to the Chebyshev polynomial filter being a relatively poor approximation to the high-energy modes of the system.
A different polynomial approximation might provide better performance, but given the simplicity of implementing tRHMC we did not consider investigating this here.
The combined PF-tRHMC filtering method performs as well as tRHMC, but again this combination is more complex to implement than tRHMC.

An important consideration for the filtering methods investigated here is choosing the filter parameters and integration step-sizes.
The most common way to tune the integration step-sizes is to choose step-sizes $h_i$ such that the molecular dynamics forces $F_i$ satisfy $F_i h_i \simeq \mathrm{constant}.$
However, when using this force balancing method to tune the step-sizes, we found that the cost function is highly dependent on the choice of filter parameters -- choose these too low or too high, and the performance compared with ordinary RHMC is only marginally better or even worse.

To mitigate the filter parameter dependence, we introduced a novel way of tuning the integration step-sizes, which we refer to as c-scale tuning.
This technique uses the characteristic scale to set the coarsest step size, then holds that fixed while tuning the next step size, with the remaining step sizes set using the force balancing method.
For both filtering methods and both lattices, we found that while employing this technique did not improve the lowest achievable cost, it did make the cost function significantly less sensitive to the filter parameters.

The advantage of this is particularly relevant for larger and more physical lattices, where generating configurations is so expensive that we cannot tune the filter parameters by brute-force within a reasonable time-frame.
When we use multiple filters, which is usually necessary for such lattices, then the benefits of c-scale tuning should prove significant. One can just choose a reasonable set of filters, calculate the forces, then tune two step-sizes to achieve near-optimal costs.

Implementing tRHMC and c-scale tuning is straightforward.
In particular, tRHMC only requires a small modification to regular RHMC code to allow for multiple rational polynomial terms.
Applying c-scale tuning only requires information about the force terms in a HMC simulation.
Thus, these techniques can be quickly deployed in order to reduce the computational cost in simulating single-flavour pseudofermions on the lattice.
Here, the lightest quark mass we considered was $m_\pi \simeq 300$ MeV.
At lighter masses, the relative benefits of single-flavour tRHMC filtering will increase.
This should prove particularly useful for the current trend of dynamical QCD+QED configurations, where the up and the down quark must be simulated separately.

\section*{Acknowledgements}
We have used BQCD~\cite{BQCD} to generate the configurations in this work.
This work was supported by resources provided by The Pawsey Supercomputing Centre with funding from the Australian Government and the Government of Western Australia,
along with resources from the National Computational Infrastructure (NCI), which is supported by the Australian Government.
The authors gratefully acknowledge the Gauss Centre for Supercomputing e.V. (www.gauss-centre.eu) for funding this project by providing computing time on the GCS Supercomputer JUQUEEN at J\"ulich Supercomputing Centre (JSC).
This investigation was supported by the Australian Research Council
under Grant Numbers FT100100005, DP120104627, DP140103067, and DP150103164.

\appendix

\section{Force terms: derivation} \label{app:forces}
\tikzsetfigurename{figure_A.}

\subsection{RHMC}
The force term for the pure RHMC action \eqref{eq:rhmc_action} takes the form
\begin{align*}
	F_{RHMC}
&= \diff{S_{RHMC}}{U} = \diff{}{U} \left( \phi^\dag R(K) \phi\right) \\
&= \diff{}{U} \left( \phi^\dag c_n \prod_{i=1}^{n} \frac{K+a_i}{K+b_i}  \phi\right) \quad \mathrm{[Using\ \eqref{eq:ratapp}]} \\
&= \phi^\dag \diff{}{U} \left( c_n \prod_{i=1}^{n} \frac{K+a_i}{K+b_i} \right) \phi \\
&= \phi^\dag \diff{}{U} \left(
	\sum_{i=1}^{n} c_n  \frac{\prod_{j=1}^{n}(-b_j+a_i)}{\prod_{j=1,j \neq i}^{n} (-b_j+b_i)} \frac{1}{K+b_i}
	\right) \phi \\
&= \phi^\dag \diff{}{U} \left(
	\sum_{i=1}^{n} \frac{r_i}{K+b_i}
	\right) \phi \\
&= \sum_{i=1}^{n} \phi^\dag [K+b_i]^{-1} \diff{K}{U} r_i [K+b_i]^{-1} \phi,
\end{align*}
where we define the real numbers
\begin{equation*}
	r_i = c_n \frac{\prod_{j=1}^{n}(-b_j+a_i)}{\prod_{j=1,j \neq i}^{n} (-b_j+b_i)}.
\end{equation*}
This form is suggestive of using a multi-shift iterative solver such as multi-shift conjugate gradient to simultaneously calculate the shifted inverses $[K+b_i]^{-1} \phi$.

\subsection{tRHMC}
Truncated RHMC \eqref{eq:trhmc_action} uses truncated sums of the rational approximation, so the force terms involved take the same form as in pure RHMC with different sum and product limits: for the term $\phi^\dag R_{k,l}(K) \phi$, we have force
\begin{align*}
	F_{tRHMC}
= \sum_{i=k+1}^{l} \phi^\dag [K+b_i]^{-1} \diff{K}{U} r_i [K+b_i]^{-1} \phi
\end{align*}
with
\begin{equation*}
	r_i = (c_n)^{\delta_{k0}} \frac{\prod_{j=k+1}^{l}(-b_j+a_i)}{\prod_{j=k+1,j \neq i}^{l} (-b_j+b_i)}
\end{equation*}
noting that the normalisation $c_n$ is only include if $k=0$.

\subsection{PF-RHMC}
In the case of PF-RHMC \eqref{eq:pfrhmc}, we have two action terms: the polynomial filter term and the correction term.

Inserting an root product expression of an $m$-th order polynomial, the polynomial filter term's force $F_1$ takes the form
\begin{align*}
	F_1
&= \phi_1^\dag \diff{P(K)}{U} \phi_1 \\
&= \phi_1^\dag \diff{}{U} \left( d_m \prod_{i=1}^{m}(K - z_i) \right) \phi_1 \\
&= \sum_{i=1}^{m} \eta_i^\dag \diff{K}{U} \chi_i,
\end{align*}
where
\begin{align*}
	\chi_i &= d_m \prod_{j=i+1}^{m} (K - z_i) \phi_1 \\
	\mathrm{and} \quad \eta_i &= \prod_{j=1}^{i-1} (K - z_i^*) \phi_1.
\end{align*}
Note that this force does not involve inverses $K^{-1}$, which leads to a low computational cost.

The correction term's force $F_2$ can be written as
\begin{align*}
	F_2
&= \phi_2^\dag \diff{}{U} \left( \frac{c_n}{d_m} \prod_{i=1}^{n} \frac{1}{K - z_i} \prod_{i=1}^{n} \frac{K+a_i}{K+b_i} \right) \phi_2 \\
&= \phi_2^\dag \diff{}{U} \left( \frac{c_n}{d_m} \frac{\prod_{i=1}^{n} (K+a_i)}{\prod_{i=1}^{m+n} (K + b_i)} \right) \phi_2 \qquad [\mathrm{defining\ } b_{n+i} = z_i] \\
&= \phi_2^\dag \diff{}{U} \left(
	\sum_{i=1}^{m+n} \frac{c_n}{d_m} \frac{\prod_{j=1}^{n}(-b_j+a_i)}{\prod_{j=1,j \neq i}^{m+n} (-b_j+b_i)} \frac{1}{K+b_i}
	\right) \phi_2 \\
&= \phi_2^\dag \diff{}{U} \left(
	\sum_{i=1}^{m+n} \frac{R_i}{K+b_i}
	\right) \phi_2 \\
&= \sum_{i=1}^{m+n} \phi_2^\dag [K+b_i]^{-1} \diff{K}{U} R_i [K+b_i]^{-1} \phi_2,
\end{align*}
where we define the complex numbers
\begin{equation*}
	R_i = \frac{c_n}{d_m} \frac{\prod_{j=1}^{n}(-b_j+a_i)}{\prod_{j=1,j \neq i}^{m+n} (-b_j+b_i)}.
\end{equation*}
The form of this force is very similar to pure RHMC, but now we have some \emph{complex} shifted inverses $[K-z_i]^{-1}$. We can calculate these using a complex multi-shift iterative solver.

\section{Force terms: data} \label{app:forces_data}
\tikzsetfigurename{figure_B.}

\subsection{\texorpdfstring{Forces on the $16^3 \times 32$ lattice}%
							{Forces on the 16x32 lattice}}

This section contains both plots (\ref{fig:force_1f}, \ref{fig:force_2f}) and tables (\ref{tbl:force_1p}, \ref{tbl:force_1t}, \ref{tbl:force_2p}, \ref{tbl:force_2t}, \ref{tbl:force_pt}) of the average and maximal forces of each fermion action terms for the actions on the $16^3 \times 32$ lattice.
All figures quoted have statistical uncertainty roughly equal to the last significant figure.
For all actions considered, the gauge action $S_G[U]$ has the force $F_G = 9.76 \mathrm{\ (avg)}$,$ 18.7 \mathrm{\ (max)}$.

\begin{table}[htb]
\centering
\caption{PF-RHMC forces on the $16^3 \times 32$ lattice. The $p=0$ entry is plain RHMC, included for reference. \label{tbl:force_1p}}
\begin{tabular}{lllll} \toprule
& \multicolumn{2}{c}{$F_1$} & \multicolumn{2}{c}{$F_2$} \\ \cmidrule(lr){2-3} \cmidrule(lr){4-5}
$p$ & avg & max & avg & max \\ \midrule
0 & -- & -- & 1.69 & 4.93 \\
4 & 1.97 & 5.78 & 1.11 & 2.97 \\
10 & 1.88 & 5.30 & 0.80 & 2.03 \\
16 & 1.82 & 5.15 & 0.65 & 1.62 \\ \bottomrule
\end{tabular}
\end{table}

\begin{table}[htb]
\centering
\caption{tRHMC forces on the $16^3 \times 32$ lattice. The $t=0$ entry is plain RHMC, included for reference.  \label{tbl:force_1t}}
\begin{tabular}{lllll} \toprule
& \multicolumn{2}{c}{$F_1$} & \multicolumn{2}{c}{$F_2$} \\ \cmidrule(lr){2-3} \cmidrule(lr){4-5}
$t$ & avg & max & avg & max \\ \midrule
0 & -- & -- & 1.69 & 4.93 \\
2 & 0.412 & 1.18 & 1.307 & 3.84 \\
3 & 0.877 & 2.52 & 0.860 & 2.58 \\
4 & 1.287 & 3.72 & 0.453 & 1.43 \\
5 & 1.522 & 4.41 & 0.211 & 0.779 \\
6 & 1.626 & 4.72 & 0.0944 & 0.452 \\
7 & 1.667 & 4.84 & 0.0458 & 0.291 \\
8 & 1.683 & 4.90 & 0.0219 & 0.182 \\ \bottomrule
\end{tabular}
\end{table}

\begin{table}[htb]
\centering
\caption{2PF-RHMC forces on the $16^3 \times 32$ lattice. \label{tbl:force_2p}}
\begin{tabular}{llllllll} \toprule
& & \multicolumn{2}{c}{$F_1$} & \multicolumn{2}{c}{$F_2$} & \multicolumn{2}{c}{$F_3$} \\ \cmidrule(lr){3-4} \cmidrule(lr){5-6} \cmidrule(lr){7-8}
$p$ & $q$ & avg & max & avg & max & avg & max \\ \midrule
4 & 10 & 1.970 & 5.78 & 1.147 & 3.08 & 0.2505 & 0.650 \\
& 16 & 1.970 & 5.78 & 1.123 & 3.01 & 0.1811 & 0.469 \\
& 20 & 1.970 & 5.78 & 1.118 & 3.00 & 0.1617 & 0.415 \\ \bottomrule
\end{tabular}
\end{table}

\begin{table}[htb]
\centering
\caption{2tRHMC forces on the $16^3 \times 32$ lattice. \label{tbl:force_2t}}
\begin{tabular}{llllllll} \toprule
& & \multicolumn{2}{c}{$F_1$} & \multicolumn{2}{c}{$F_2$} & \multicolumn{2}{c}{$F_3$} \\ \cmidrule(lr){3-4} \cmidrule(lr){5-6} \cmidrule(lr){7-8}
$t$ & $t'$ & avg & max & avg & max & avg & max \\ \midrule
4 & 5 & 1.287 & 3.72 & 0.2524 & 0.758 & 0.2103 & 0.775 \\
& 6 & 1.287 & 3.72 & 0.3702 & 1.126 & 0.0947 & 0.454 \\
& 7 & 1.287 & 3.72 & 0.4193 & 1.287 & 0.0447 & 0.286 \\
& 8 & 1.287 & 3.72 & 0.4387 & 1.358 & 0.0212 & 0.180 \\
5 & 7 & 1.522 & 4.41 & 0.1717 & 0.563 & 0.0443 & 0.284 \\
& 8 & 1.522 & 4.41 & 0.1931 & 0.653 & 0.0212 & 0.173 \\
& 10 & 1.522 & 4.40 & 0.2060 & 0.730 & 0.0045 & 0.061 \\ \bottomrule
\end{tabular}
\end{table}

\begin{table}[htb]
\centering
\caption{PF-tRHMC forces on the $16^3 \times 32$ lattice.\label{tbl:force_pt}}
\begin{tabular}{llllllll} \toprule
& & \multicolumn{2}{c}{$F_1$} & \multicolumn{2}{c}{$F_2$} & \multicolumn{2}{c}{$F_3$} \\ \cmidrule(lr){3-4} \cmidrule(lr){5-6} \cmidrule(lr){7-8}
$p$ & $t$ & avg & max & avg & max & avg & max \\ \midrule
4 & 4 & 1.274 & 3.70 & 0.2010 & 0.542 & 0.4540 & 1.433 \\
& 5 & 1.582 & 4.66 & 0.5680 & 1.528 & 0.2094 & 0.767 \\
& 6 & 1.806 & 5.31 & 0.8798 & 2.366 & 0.0946 & 0.453 \\
& 7 & 1.914 & 5.62 & 1.028 & 2.764 & 0.0455 & 0.291 \\ \bottomrule
\end{tabular}
\end{table}

\FloatBarrier

\subsection{\texorpdfstring{Forces on the $24^3 \times 48$ lattice}%
							{Forces on the 24x48 lattice}}

This section contains tables (\ref{tbl:force_1t_24x48}, \ref{tbl:force_2t_24x48}) of the average and maximal forces of each fermion action term for the fermion actions on the $24^3 \times 48$ lattice.
All figures quoted have statistical uncertainty roughly equal to the last significant figure.

For all actions considered: the gauge action $S_G[U]$ has the force $F_G = 9.28 \mathrm{\ (avg)}, 18.0 \mathrm{\ (max)}$, and the determinant term $S_{det}$ has the force $F_{det} = 0.299 \mathrm{\ (avg)}, 0.825 \mathrm{\ (max)}$.

\begin{table}[htb]
\centering
\caption{tRHMC forces on the $24^3 \times 48$ lattice. The $t=0$ entry is plain RHMC, included for reference.  \label{tbl:force_1t_24x48}}
\begin{tabular}{lllll} \toprule
& \multicolumn{2}{c}{$F_1$} & \multicolumn{2}{c}{$F_2$} \\ \cmidrule(lr){2-3} \cmidrule(lr){4-5}
$t$ & avg & max & avg & max \\ \midrule
0 & -- & -- & 2.64 & 9.19 \\
6 & 2.3798 & 7.80 & 0.335 & 3.8 \\
8 & 2.5872 & 8.77 & 0.0872 & 2.2 \\
10 & 2.6269 & 9.02 & 0.0305 & 1.26 \\
12 & 2.6352 & 9.12 & 0.0114 & 0.74 \\ \bottomrule
\end{tabular}
\end{table}

\begin{table}[htb]
\centering
\caption{2tRHMC forces on the $24^3 \times 48$ lattice. \label{tbl:force_2t_24x48}}
\begin{tabular}{llllllll} \toprule
& & \multicolumn{2}{c}{$F_1$} & \multicolumn{2}{c}{$F_2$} & \multicolumn{2}{c}{$F_3$} \\ \cmidrule(lr){3-4} \cmidrule(lr){5-6} \cmidrule(lr){7-8}
$t$ & $t'$ & avg & max & avg & max & avg & max \\ \midrule
4 & 8 & 1.6196 & 5.10 & 1.0687 & 4.71 & 0.0901 & 2.19 \\
& 10 & 1.6197 & 5.11 & 1.1172 & 5.28 & 0.0304 & 1.41 \\
& 12 & 1.6198 & 5.11 & 1.1274 & 5.43 & 0.0106 & 0.68 \\ \bottomrule
\end{tabular}
\end{table}

\FloatBarrier

\section{Molecular dynamics data} \FloatBarrier
\tikzsetfigurename{figure_B.}

\subsection{\texorpdfstring{$16^3 \times 32$ lattice}%
							{16x32 lattice}}

\begin{table}[!htbp]
\centering
\caption{PF-RHMC configurations on the $16^3 \times 32$ lattice, using force balancing \label{tbl:pfrhmc}}

\begin{tabular}{@{}llllll@{}} \toprule
$p$ & $n_2$ & $n_1$ & $n_0$ & $P_{acc}$ & $N_{\mathrm{traj}}$ \\ \midrule
4 & 10 & 18 & 480 & 0.68(1) & 2000 \\
10 & 10 & 26 & 480 & 0.70(1) & 2000 \\
16 & 11 & 35 & 480 & 0.75(1) & 2000 \\ \bottomrule
\end{tabular}
\end{table}

\begin{table}[!htbp]
\centering
\caption{C-scale tuned configurations for PF-RHMC on the $16^3 \times 32$ lattice \label{tbl:pfrhmc_cs}}

\begin{tabular}{@{}llllll@{}} \toprule
$p$ & $n_2$ & $n_1$ & $n_0$ & $P_{acc}$ & $N_{\mathrm{traj}}$ \\ \midrule
4 & 10 & 30 & 480 & 0.71(1) & 2000 \\
10 & 10 & 30 & 480 & 0.72(1) & 2000 \\
16 & 11 & 35 & 480 & 0.75(1) & 2000 \\ \bottomrule
\end{tabular}%
\end{table}

\begin{table}[!htbp]
\centering
\caption{tRHMC configurations on the $16^3 \times 32$ lattice, using force balancing\label{tbl:trhmc}}

\begin{tabular}{@{}llllll@{}} \toprule
$t$ & $n_2$ & $n_1$ & $n_0$ & $P_{acc}$ & $N_{\mathrm{traj}}$ \\ \midrule
2 & 15 & 5 & 480 & 0.68(1) & 2000 \\
3 & 10 & 10 & 480 & 0.72(1) & 2000 \\
4 & 8 & 20 & 480 & 0.67(1) & 2000 \\
5 & 6 & 35 & 480 & 0.65(1) & 2000 \\
6 & 5 & 60 & 480 & 0.73(1) & 2000 \\
7 & 4 & 70 & 480 & 0.70(1) & 2000 \\
8 & 3 & 85 & 480 & 0.68(1) & 2000 \\ \bottomrule
\end{tabular}
\end{table}

\begin{table}[!htbp]
\centering
\caption{C-scale tuned configurations for tRHMC on the $16^3 \times 32$ lattice \label{tbl:trhmc_cs}}

\begin{tabular}{@{}llllll@{}} \toprule
$t$ & $n_2$ & $n_1$ & $n_0$ & $P_{acc}$ & $N_{\mathrm{traj}}$ \\ \midrule
2 & 11 & 25 & 480 & 0.71(1) & 2000 \\
3 & 10 & 25 & 480 & 0.76(1) & 2000 \\
4 & 8 & 20 & 480 & 0.67(1) & 1500 \\
5 & 6 & 30 & 480 & 0.70(1) & 2000 \\
6 & 5 & 15 & 480 & 0.71(1) & 2000 \\
7 & 4 & 15 & 480 & 0.67(1) & 2000 \\
8 & 3 & 25 & 480 & 0.65(1) & 2000 \\ \bottomrule
\end{tabular}
\end{table}

\begin{table}[!htbp]
\centering
\caption{2PF-RHMC configurations on the $16^3 \times 32$ lattice, using force balancing \label{tbl:2pfrhmc_fdt}}

\begin{tabular}{@{}llllllll@{}} \toprule
$p$ & $q$ & $n_3$ & $n_2$ & $n_1$ & $n_0$ & $P_{acc}$ & $N_{\mathrm{traj}}$ \\ \midrule
4 & 10 & 7 & 33 & 62 & 480 & 0.68(1) & 2000 \\
& 16 & 8 & 51 & 98 & 480 & 0.66(1) & 2000 \\
& 20 & 9 & 65 & 125 & 480 & 0.70(1) & 2000 \\ \bottomrule
\end{tabular}
\end{table}

\begin{table}[!htbp]
\centering
\caption{C-scale tuned 2PF-RHMC configurations on the $16^3 \times 32$ lattice \label{tbl:2pfrhmc}}

\begin{tabular}{@{}llllllll@{}} \toprule
$p$ & $q$ & $n_3$ & $n_2$ & $n_1$ & $n_0$ & $P_{acc}$ & $N_{\mathrm{traj}}$ \\ \midrule
4 & 10 & 7 & 33 & 62 & 480 & 0.68(1) & 2000 \\
& 16 & 8 & 15 & 27 & 480 & 0.65(1) & 2000 \\
& 20 & 9 & 20 & 38 & 480 & 0.72(1) & 2000 \\ \bottomrule
\end{tabular}
\end{table}

\begin{table}[!htbp]
\centering
\caption{2tRHMC configurations on the $16^3 \times 32$ lattice, using force balancing\label{tbl:2trhmc_fdt}}

\begin{tabular}{@{}llllllll@{}} \toprule
$t$ & $t'$ & $n_3$ & $n_2$ & $n_1$ & $n_0$ & $P_{acc}$ & $N_{\mathrm{traj}}$ \\ \midrule
4 & 5 & 6 & 6 & 29 & 480 & 0.69(1) & 2000 \\
& 6 & 5 & 12 & 40 & 480 & 0.72(1) & 2000 \\
& 7 & 4 & 17 & 50 & 480 & 0.71(1) & 2000 \\
& 8 & 3 & 25 & 70 & 480 & 0.66(1) & 2000 \\
5 & 7 & 4 & 8 & 64 & 480 & 0.67(1) & 2000 \\
& 8 & 3 & 10 & 70 & 480 & 0.66(1) & 2000 \\
& 10 & 2 & 22 & 129 & 480 & 0.77(1) & 2000 \\ \bottomrule
\end{tabular}
\end{table}

\begin{table}[!htbp]
\centering
\caption{C-scale tuned 2tRHMC configurations on the $16^3 \times 32$ lattice \label{tbl:2trhmc}}

\begin{tabular}{@{}llllllll@{}} \toprule
$t$ & $t'$ & $n_3$ & $n_2$ & $n_1$ & $n_0$ & $P_{acc}$ & $N_{\mathrm{traj}}$  \\ \midrule
4 & 5 & 6 & 6 & 29 & 480 & 0.69(1) & 2000 \\
& 6 & 5 & 12 & 40 & 480 & 0.72(1) & 2000 \\
& 7 & 4 & 12 & 35 & 480 & 0.69(1) & 2000 \\
& 8 & 3 & 10 & 30 & 480 & 0.67(1) & 2000 \\
5 & 7 & 4 & 6 & 47 & 480 & 0.71(1) & 2000 \\
& 8 & 3 & 5 & 35 & 480 & 0.63(1) & 2000 \\
& 10 & 2 & 5 & 30 & 480 & 0.68(1) & 2000 \\ \bottomrule
\end{tabular}
\end{table}

\begin{table}[!htbp]
\centering
\caption{PF-tRHMC configurations on the $16^3 \times 32$ lattice, using force balancing \label{tbl:pftrhmc_fdt}}

\begin{tabular}{@{}llllllll@{}} \toprule
$p$ & $t$ & $n_3$ & $n_2$ & $n_1$ & $n_0$ & $P_{acc}$ & $N_{\mathrm{traj}}$  \\ \midrule
4 & 4 & 15 & 6 & 39 & 480 & 0.82(1) & 2000 \\
& 5 & 6 & 12 & 36 & 480 & 0.75(1) & 2000 \\
& 6 & 5 & 26 & 59 & 480 & 0.75(1) & 2000 \\
& 7 & 4 & 38 & 79 & 480 & 0.69(1) & 2000 \\ \bottomrule
\end{tabular}
\end{table}

\begin{table}[!htbp]
\centering
\caption{C-scale tuned PF-tRHMC configurations on the $16^3 \times 32$ lattice \label{tbl:pftrhmc}}

\begin{tabular}{@{}llllllll@{}} \toprule
$p$ & $t$ & $n_3$ & $n_2$ & $n_1$ & $n_0$ & $P_{acc}$ & $N_{\mathrm{traj}}$  \\ \midrule
4 & 4 & 8 & 8 & 42 & 480 & 0.73(1) & 2000 \\
& 5 & 6 & 10 & 30 & 480 & 0.69(1) & 2000 \\
& 6 & 5 & 12 & 27 & 480 & 0.69(1) & 2000 \\
& 7 & 4 & 16 & 32 & 480 & 0.70(1) & 2000 \\ \bottomrule
\end{tabular}
\end{table}

\FloatBarrier
\subsection{\texorpdfstring{$24^3 \times 48$ lattice}%
							{24x48 lattice}}

\begin{table}[!htbp]
\centering
\caption{C-scale tuned configs for tRHMC on the $24^3 \times 48$ lattice \label{tbl:24x48_trhmc_cs}}

\begin{tabular}{@{}llllll@{}} \toprule
$t$ & $n_2$ & $n_1$ & $n_0$ & $P_{acc}$ & $N_{\mathrm{traj}}$ \\ \midrule
6 & 20 & 30 & 480 & 0.71(4) & 100 \\
8 & 16 & 25 & 480 & 0.69(5) & 100 \\
10 & 12 & 25 & 480 & 0.73(4) & 100 \\
12 & 10 & 25 & 480 & 0.66(5) & 100 \\ \bottomrule
\end{tabular}
\end{table}

\begin{table}[!htbp]
\centering
\caption{C-scale tuned configs for 2tRHMC on the $24^3 \times 48$ lattice \label{tbl:24x48_2trhmc_cs}}

\begin{tabular}{@{}llllllll@{}} \toprule
$t$ & $t'$ & $n_3$ & $n_2$ & $n_1$ & $n_0$ & $P_{acc}$ & $N_{\mathrm{traj}}$ \\ \midrule
4 & 8 & 15 & 30 & 32 & 480 & 0.70(4) & 100 \\
& 10 & 12 & 35 & 34 & 480 & 0.66(5) & 100 \\
& 12 & 11 & 25 & 23 & 480 & 0.65(5) & 100 \\ \bottomrule
\end{tabular}
\end{table}

\FloatBarrier

\section{Auxiliary data}

\subsection{Additional tables}

\begin{table}[htb]
\centering
\caption{The coefficients in \eqref{eq:ratapp} of the 20th order Zolotarev approximation, optimised for the range $[5 \times 10^{-5}, 3]$. \label{tbl:zolo_eg}}
\begin{tabular}{ccc} \toprule
$k$ & $a_k$ & $b_k $ \\ \midrule
1 & 95.6316331717 & 21.8515265744 \\
2 & 8.38850349943 & 3.86925533340 \\
3 & 1.93840206679 & 1.01168874377 \\
4 & 0.539371178581 & 0.290831421623 \\
5 & 0.157775824154 & $0.858763902980\times 10^{-1}$ \\
6 & $0.468260410763\times 10^{-1}$ & $0.255579785943\times 10^{-1}$ \\
7 & $0.139571812255\times 10^{-1}$ & $0.762422486830\times 10^{-2}$ \\
8 & $0.416545933719\times 10^{-2}$ & $0.227597709395\times 10^{-2}$ \\
9 & $0.124363649128\times 10^{-2}$ & $0.679563677515\times 10^{-3}$ \\
10 & $0.371340852888\times 10^{-3}$ & $0.202916856414\times 10^{-3}$ \\
11 & $0.110882844025\times 10^{-3}$ & $0.605912276150\times 10^{-4}$ \\
12 & $0.331094772774\times 10^{-4}$ & $0.180921019109\times 10^{-4}$ \\
13 & $0.988586317473\times 10^{-5}$ & $0.540156470610\times 10^{-5}$ \\
14 & $0.295111943023\times 10^{-5}$ & $0.161207322288\times 10^{-5}$ \\
15 & $0.880351240756\times 10^{-6}$ & $0.480501824158\times 10^{-6}$ \\
16 & $0.262004468388\times 10^{-6}$ & $0.142607387932\times 10^{-6}$ \\
17 & $0.773643984516\times 10^{-7}$ & $0.417152400208\times 10^{-7}$ \\
18 & $0.222400404411\times 10^{-7}$ & $0.116074993902\times 10^{-7}$ \\
19 & $0.581507245934\times 10^{-8}$ & $0.268224176045\times 10^{-8}$ \\
20 & $0.102967547829\times 10^{-8}$ & $0.235276800436\times 10^{-9}$ \\
\midrule
\multicolumn{3}{c}{
$c_n =  6.41196938508 \times 10^{-2}$
} \\ \bottomrule
\end{tabular}
\end{table}

\FloatBarrier

\section*{References}
\bibliographystyle{elsarticle-num}
\bibliography{references}

\end{document}